\documentclass[%
 aps,
superscriptaddress,
groupedaddress,
showpacs,
 amsmath,amssymb,amsthm,mathrsfs,appendix
prl,
twocolumn,
floatfix
]{revtex4-1}

\usepackage{graphicx}  
\usepackage{color}
\usepackage{dcolumn}   
\usepackage{bm}        
\usepackage{comment}
\usepackage{subfigure}

\usepackage{verbatim}

\graphicspath{{Pictures/}}

\begin{document}

\title{Extinction in complex communities as driven by adaptive dynamics}

\author{Vu A.T. Nguyen}
\affiliation{University of Notre Dame, South Bend, IN}
\author{Dervis Can Vural}
\affiliation{University of Notre Dame, South Bend, IN}
\affiliation{Correspondence to dvural@nd.edu}
\date{\today}

\begin{abstract}
In a complex community, species continuously adapt to each other. On rare occasions, the adaptation of a species can lead to the extinction of others, and even its own. ``Adaptive dynamics'' is the standard mathematical framework to describe evolutionary changes in community interactions, and in particular, predict adaptation driven extinction. Unfortunately, most authors implement the equations of adaptive dynamics through computer simulations, that require assuming a large number of questionable parameters and fitness functions. In this study we present analytical solutions to adaptive dynamics equations, thereby clarifying how outcomes depend on any computational input. We develop general formulas that predict equilibrium abundances over evolutionary time scales. Additionally, we predict which species will go extinct next, and when this will happen.
\end{abstract}
\maketitle

\section{Introduction}

For many ecological communities, the changes in population abundances due to interspecies interactions are much faster than the evolutionary changes in the traits that mediate these interactions. This affords us the theoretical convenience of dealing with population dynamics equations with unchanging interaction parameters \cite{turchin2003complex}. However, to describe the population dynamics of interacting communities over long periods, or of those that evolve rapidly, one must also take into account how species coadapt and interact differently over time \cite{dawkins1979arms,weitz2005coevolutionary,maor2005arms,stahl2000plant, dietl2002fossil,vural2015organization}. To illustrate, when a pathogen crosses a species barrier, a previously zero interaction parameter between it and its host becomes finite. Conversely, when a pathogen (perhaps the same pathogen) kills all but resistant individuals, a previously-finite interaction parameter drops to zero.

Amending population dynamics equations to include such adaptive changes to interspecies interactions is referred as ``adaptive dynamics'' (AD) \cite{schaffer1978homage,rosenzweig1978homage,rosenzweig1987red,hochberg1995refuge,dieckmann1995evolutionary,dieckmann1996dynamical,marrow1996evolutionary,gavrilets1997coevolutionary,abrams1999adaptive,abrams2000evolution,gandon2008host,zu2016evolutionary,loeuille2018multidimensionality,lion2018theoretical,lehtinen2019coevolution,cortez2020destabilizing}. We should caution that AD is applicable only to asexual, well-mixed communities, in the low mutation rate limit.

As with much of classical population dynamics, AD typically focuses on demonstrating the stability of communities -- even in the face of perpetual evolutionary arms races \cite{weitz2005coevolutionary,zu2016evolutionary,cortez2017effects,lehtinen2019coevolution,cortez2020destabilizing}. However, adaptive changes in interspecies interactions can occasionally lead to catastrophic displacements in equilibrium abundances, and even extinction, as was suggested theoretically \cite{marrow1996evolutionary,matsuda1994timid,matsuda1994runaway,gyllenberg2001necessary,leimar2002evolutionary,parvinen2005evolutionary,parvinen2010adaptive,parvinen2013self,boldin2016evolutionary} and empirically \cite{muir1999possible,conover2002sustaining,fiegna2003competitive,howard2004transgenic,olsen2004maturation,j2005can}. This phenomenon is referred as ``evolutionary suicide''.
Analytical studies of adaptive extinction have so far been limited to only single species mutating \cite{matsuda1994timid,matsuda1994runaway,gyllenberg2001necessary,parvinen2005evolutionary,parvinen2013self,boldin2016evolutionary} or coevolving two-species communities \cite{marrow1996evolutionary,parvinen2010adaptive,leimar2002evolutionary}.

Prey-predator, host-parasite, plant-herbivore coevolution, as well as suicidal over-hunting, is standard textbook material. However, in a complex community, arms races need not be an exclusively two-species phenomenon. The equilibrium abundances of two species may change only marginally as they coevolve, while still causing a chain of events that lead to the extinction of other, possibly even far removed, species. Furthermore, in a complex community, evolutionary arms races need not take place between two species privately and end when one goes extinct. Arms races could involve so many species and span such long times that speciation and extinction events can be viewed as ordinary background events that merely decorate the Hutchinsonian evolutionary stage.

In this paper we are primarily concerned with calculating the rate of extinction in a complex community, as driven by the parallel coevolution of its members. We will be interested in predicting which species will next go extinct, when this will happen, and how fast a community will loose its members to complex evolutionary arms races. While doing so, we will also obtain how equilibrium abundances and interaction mediating traits evolve.

As general, widely applicable, and conceptually insightful AD models are, in practice, questions of this kind can presently be addressed only by elaborate computer simulations that require a specific choice of initial conditions for the species abundances and interaction structure. Furthermore, these initial conditions must be accompanied by an army of functions that relate evolving traits to evolving interspecies interaction parameters. These functions are parameterized by an army of constants, typically randomly chosen according to specific probability distributions, which themselves are parameterized by arbitrarily chosen constants.

For example, \cite{ito2007new} discovered that if there is frequency-dependent disruptive selection in one trait and weak directional selection in another, then the community can enter a cycle of recurrent adaptive radiations and extinctions. Although this provides a mechanism for continuous introduction and removal of species it requires very particular interaction functions that allow trait branching. Likewise, \cite{johansson2009evolutionary} investigated the consequences of extinctions in ecological communities, and found evolutionary keystone species that can switch the community from one evolutionary domain of attraction to another. Although their work carefully describes the convergence landscape, due to the combinatorially large possible number of trajectories it is extremely difficult to gain general insights into how fast a large, complex community would decline under adaptive dynamics. Other theoretical studies that examine extinction-speciation dynamics suffer from the similar difficulties, of having to assume highly specific parameter sets \cite{kisdi2002red,dercole2003remarks,ito2014evolutionary}. 

\begin{figure*}
\centering
\includegraphics[width=\linewidth]{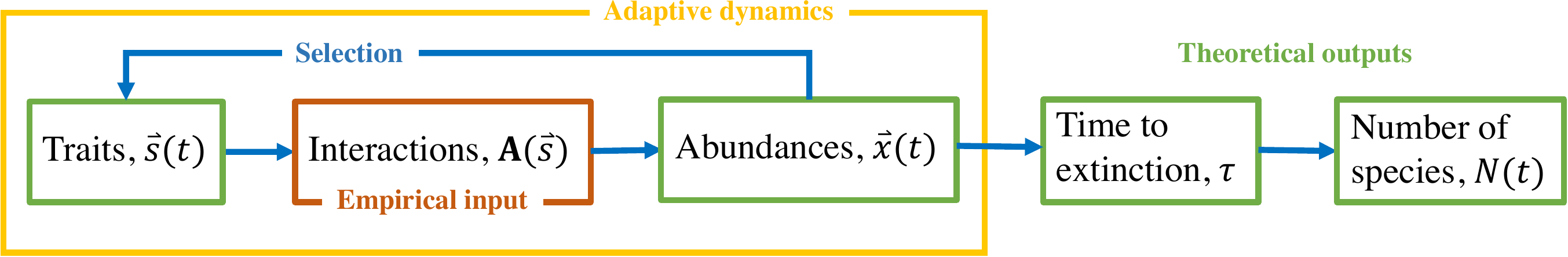}
\caption{{\bf Schematics of our approach}. Here we analytically solve canonical adaptive dynamics equations (yellow), where interaction mediating traits evolve as to climb up a fitness landscape, as defined according to Lotka-Volterra-type equations. We take the trait dependence of interspecies interactions ${\bf A}(\vec{s})$ as given (red), and from here, calculate (green), (1) how equilibrium abundances shift, $\vec{x}(t)$; (2) the time $\tau$ it takes for this shift to lead to a single extinction and (3) the species that gets extinct the soonest; (4) the number of species as a function of time $N(t)$, as many such extinctions accumulate; and (5) the evolution of the traits $\vec{s}(t)$ while all the above is happening. The blue arrows indicate causal links, and also show which quantity is used to calculate which other. }
\label{fig:schematics}
\end{figure*}

Given the high dimensionality of initial conditions and model parameters, and given the sensitivity of the outcomes to the precise values of some of these parameters, numerical simulations provide little general insight into the macroscopic consequences of adaptive dynamics for short, medium, and long time scales. For short time scales, we should be able to predict how species abundances change upon evolving interspecies interactions. For medium time scales we should be able to predict how often a species will go extinct upon an accumulation of such adaptations; and for long time scales, we should be able to predict if and when an entire community would accumulate enough adaptations to annihilate itself. Furthermore, we should be able to make these predictions for any community, without having to assume functions, parameters, or probability distributions that are not only highly specific and questionable, but also highly consequential. Lastly, since it is practically impossible to measure an interaction matrix in entirety, we would like to know which interaction parameters suffice to make ``reasonably accurate'' predictions, and empirically measure only those.

%

Here we take a step towards addressing the above questions by deriving accurate and general analytical solutions to adaptive dynamics equations. Since our results are analytical, they allow us to see clearly how population trajectories and extinction times depend on model inputs, without necessarily assuming what they are. 

Our work consists of two parts. In part one, we assume that the interspecies interactions $A_{ij}$ change in time according to ${\bf A}(t)={\bf H_0}+{\bf H_1}t+{\bf H_2}t^2+\ldots$ and calculate the equilibrium abundances and extinction times of species in terms of ${\bf H_0,H_1,H_2,\ldots}$ If $A_{ij}(t)$ is measured empirically at multiple time points, one can fit a line or polynomial and from there extrapolate extinction times. 

Despite being a useful tool for extrapolation, the result of part one is really a stepping stone. Its true strength will become apparent in part two of our paper, where we derive what the ${\bf H}$'s are starting with the canonical equations for adaptive dynamics, which take into account the coupling between evolving traits and interaction matrix elements that depend on these traits. 

Equipped with the results of both part one and two, one could now measure how the interactions ${\bf A}(\vec{s})$ depend on traits $\vec{s}=\{s_1, s_2,...\}$, fit a line to each, and then feed these slopes into our analytical formulas to get future abundances $\vec{x}(t)$, trait evolution $\vec{s}(t)$ and extinction times $\tau$.

After we analytically obtain $\vec{x}(t)$, $\vec{s}(t)$ and $\tau$, we apply them to specific AD models which we gather from the literature. We compare our general analytical formulas with the numerical solutions of three specific models (which, as is customary, we run with a large number of made-up parameter values and ad-hoc functional forms): One where ${\bf A}(t)$ is a random matrix, where each matrix element changes at its own random constant rate, one where ${\bf A}(\vec{s})$ depends on the difference between two evolving traits, as in \cite{schaffer1978homage,rosenzweig1978homage,rosenzweig1987red}, and one where ${\bf A}(\vec{s})$ has a Gaussian dependence on evolving traits, as in \cite{marrow1992coevolution,marrow1996evolutionary}. 

Throughout, we verify our analytical results with deterministic (mean-field) simulations, as well as stochastic simulations (cf. Numerical Methods), and find very good agreement.

We conclude our introduction with an exposition of canonical adaptive dynamics, as much as it relates to the present work. In classical population dynamics, the growth rate (fitness) $f_i(\vec{x})$ of the abundance $x_i$ of species $i$ depends on the abundances of other species in the community, $\vec{x}=\{x_1,x_2,...\}$. That is, $dx_i/dt= f_i(\vec{x})x_i$. 

There are many population dynamics models with varying degrees of complexity and realism \cite{turchin2003complex}, and adaptive dynamics can be imposed on any of them. Here we will follow the original AD formalism which assumes the simple Lotka-Volterra (LV) fitness  $f_i = r_i+\sum_j A_{ij}x_j$, where $A_{ij}$ quantifies how the $j^\mathrm{th}$ species affects the $i^\mathrm{th}$ species, and $r_i$ the growth rate of the species in isolation. The LV model is a rather crude representation of reality, however under certain biotic and abiotic conditions, LV does overlap with more realistic models. These conditions are well established and well understood (cf. \cite{odwyer} and references therein). Furthermore, due to its analytical simplicity, LV also forms the basis of most empiric measurements of interspecies interactions \cite{delmas2019analysing}. 
 
While standard LV equations keep $A_{ij}$ unchanging, AD considers parameters $A_{ij}(\vec{s})$ that depend on some traits $\vec{s}=\{s_1,s_2,\ldots\}$. The specific functional form of $A_{ij}(\vec{s})$ depends on the ecological details of the system of interest \cite{schaffer1978homage,rosenzweig1978homage,rosenzweig1987red,hochberg1995refuge,dieckmann1995evolutionary,dieckmann1996dynamical,marrow1996evolutionary,gavrilets1997coevolutionary,abrams1999adaptive,abrams2000evolution,gandon2008host,zu2016evolutionary,loeuille2018multidimensionality,lion2018theoretical,lehtinen2019coevolution,cortez2020destabilizing}. However the defining aspect of all AD models is that novel strains with differing traits (and therefore different interaction matrix elements) will be introduced, and subsequently, LV equations will govern the fate of these strains, thereby playing the role of natural selection.

It was shown in \cite{dieckmann1996dynamical} that in the rare mutation / fast selection limit, the traits governing the interspecies interactions evolve, on average, according to the so called canonical equation,
\begin{equation}
    \frac{ds_{k\alpha}}{dt} = \frac{1}{2}\mu_{k\alpha}\sigma_{k\alpha}^2 \frac{x_k}{\lambda} \frac{\partial f_k}{\partial s_{k\alpha}}\label{canonical}
\end{equation}
where $s_{k\alpha}$ is the $\alpha^\mathrm{th}$ trait of species $k$, $\mu_{k\alpha}$ and $\sigma^2_{k\alpha}$ are the mutation rate and variance of mutation strength for $s_{k\alpha}$, $\lambda$ is the abundance corresponding to a single individual so that $x_k/\lambda$ represents the number of individuals, and $f_k$ is the LV fitness defined above. This equation makes intuitive sense: traits change at a rate proportional to how much they increase fitness and also proportional to the rate at which a population can generate variation. This equation constitutes the starting point of the present work.

\section{Results}
Setting $dx_i/dt\to0$ with $x_i\neq0$ in the LV equations, gives the coexistent equilibrium condition:
\begin{equation}
    \vec{x}=-{\bf A^{-1}}\vec{r}.
    \label{eq:fixedpoint}
\end{equation}
from which we qualitatively see that as small changes accumulate in ${\bf A}$, the equilibrium abundances $\vec{x}$ will move along some trajectory, which occasionally might cross zero. Our goal is to determine $\vec{x}(t)$, the components $i$ of $\vec{x}$ that crosses zero , and the times $\tau_i$ for which $x_i(\tau_i)=0$.

\subsection{Implicit adaptive dynamics}
According to AD, as traits $\vec{s}$ climb up a fitness landscape, the interactions $A_{ij}(\vec{s})$ that depend on them also change, leading to a difference in equilibrium abundances. 
Suppose we somehow know $A_{ij}(t)$, in the form 
\begin{equation}
{\bf A}(t) = {\bf H_0} + {\bf H_1} t + {\bf H_2} t^2 + \dots\label{A}
\end{equation}
where the ${\bf H}$ matrices could have been obtained experimentally, by measuring ${\bf A}(t)$ at few time points and fitting a line or polynomial; or theoretically, by modeling the evolution of $\vec{s}(t)$ and knowing ${\bf A}(\vec{s})$. 

\begin{figure}
\centering
\includegraphics[width=\linewidth]{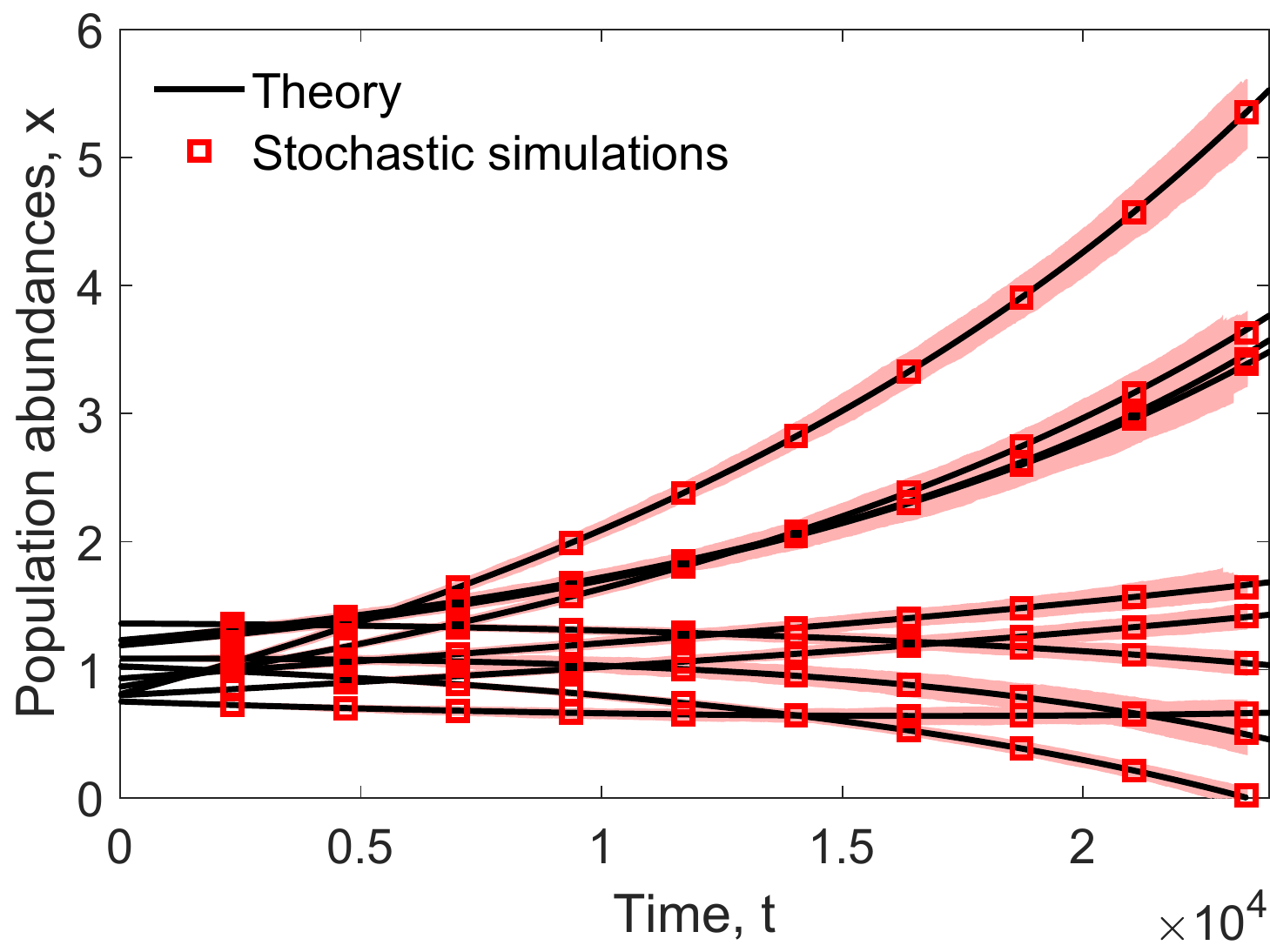}
\caption{{\bf Equilibrium abundances $\vec{x}(t)$ in a community whose interactions are changing at a fixed rate}. Our analytic formula (Eqn.\ref{exactxwithU1}) correctly tracks each species abundance over time, and can also be used to predict an extinction. We sampled $1000$ stochastic trajectories to find the min and max bounds shown by the red shaded region using $\zeta=0.1$ (cf. Numerical Methods) for a community with $N=10$ species.}
\label{fig:const_xt}
\end{figure}

\begin{figure}
\centering
\includegraphics[width=\linewidth]{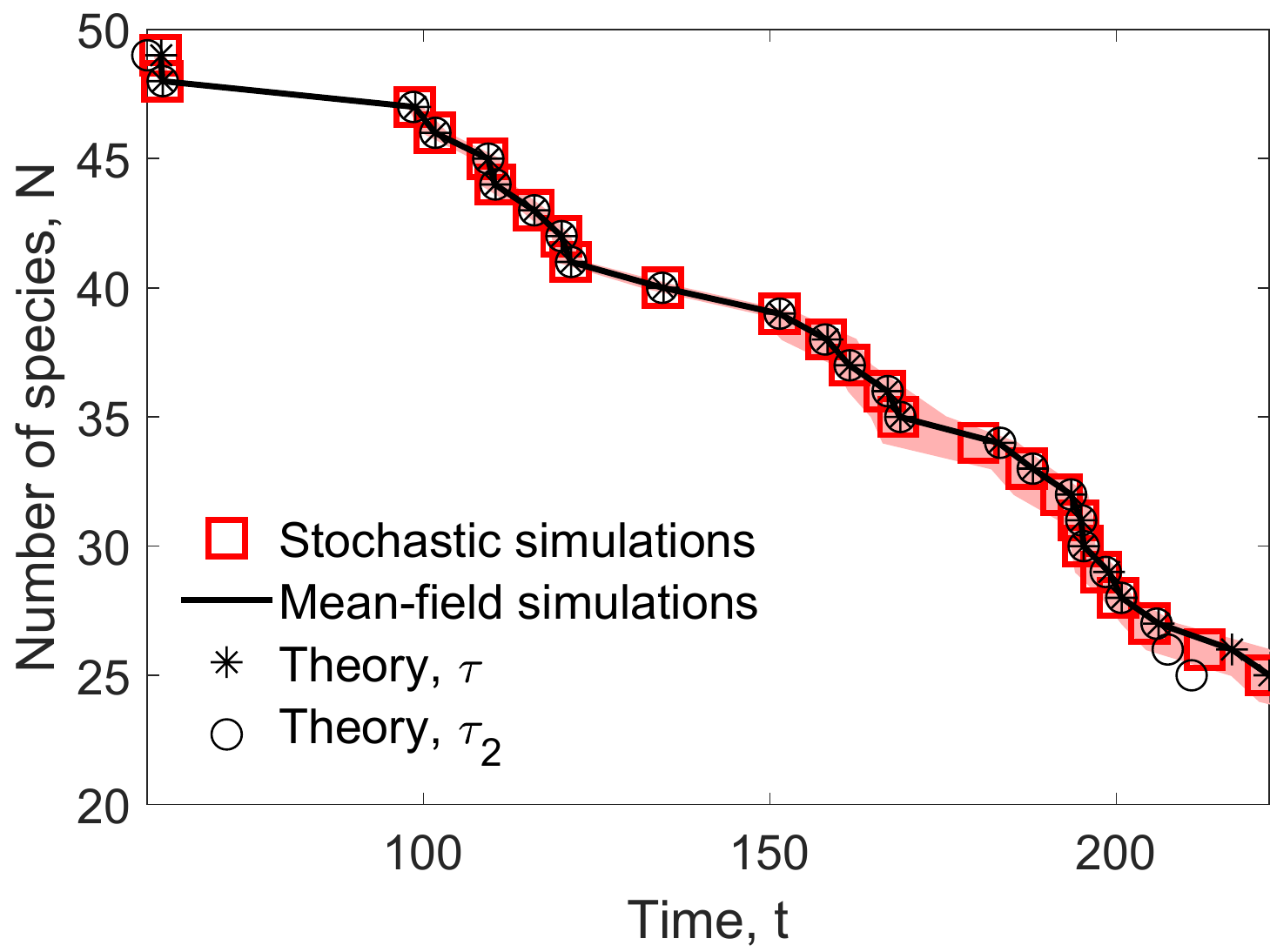}
\caption{{\bf Decline of a community as driven by species coadapting}. Here we compare the extant number of species in a coevolving community where random matrix elements of {\bf A} are mutated incrementally (stochastic simulations) and where all matrix elements change all together, continuously (mean-field simulations), to analytical theory Eqn.\ref{exactxwithU1} and the second-order approximation Eqn.\ref{tauconst}. The shaded red regions contain all ($5000$) simulated trajectories using the minimum and maximum extinction times at each community size. The initial $N=50$ community has an interaction density $\rho=1.0$ and randomly selects $50\%$ $(\zeta=0.5)$ of interaction elements to change at each time-step (cf. Numerical Methods).}
\label{fig:MC_Nt}
\end{figure}

Of course, in reality, the interaction matrix elements will not change all together continuously (mean-field dynamics). Instead, small discrete changes to randomly selected matrix elements will accumulate (stochastic dynamics). Of course, when mutation strength is small, over long periods the two approaches gives identical results (as shown in Fig.\ref{fig:const_xt} and Fig.\ref{fig:MC_Nt}). Thus, we proceed with analysis based off of a mean-field description of Eqn.\ref{A}.



We find how abundances evolve by expanding $\vec{x}(t) = \vec{\epsilon}_0 + \vec{\epsilon}_1 t + \vec{\epsilon}_2 t^2 + \ldots$, inserting this and Eqn.\ref{A} into Eqn.\ref{eq:fixedpoint}, and match the coefficients of every power of $t$ (for details, see Appendix \ref{app:general_approximation}),
\begin{equation}
    \vec{\epsilon}_m = -{\bf A_0^{-1}}\sum_{k=1}^{m} {\bf H}_{k}\vec{\epsilon}_{m-k}.\label{implicit}
\end{equation}
Lastly, we evaluate all $\vec{\epsilon}$'s and plug them into the definition of $\vec{x}(t)$
\begin{align}
\vec{x}(t) = \vec{x}_0&+[{\bf U_1}t+({\bf U_1^2+U_2})t^2+\label{x(t)}\\
  &+({\bf U_1^3}+{\bf U_1U_2+U_2U_1+U_3})t^3+\ldots]\vec{x}_0\nonumber
\end{align}
where ${\bf U_k\equiv-A_0^{-1}H_k}$. The pattern here is that the coefficient of the $m^\mathrm{th}$ power of time consists of all ${\bf U}$'s that sum up to $m$, in every possible order.  For example, the coefficient of $t^4$ is a sum of $\bf U_1^4$, $\bf U_1^2U_2$, $\bf U_2U_1^2$, $\bf U_1U_2U_1$, $\bf U_2^2$, $\bf U_1U_3$, $\bf U_3U_1$ and $\bf U_4$.

Note that Eqn.\ref{x(t)} is an infinite series and depending on the nature of the original interaction matrix ${\bf A_0}$ might have a finite radius of convergence, which would prohibit us from knowing $\vec{x}(t)$ beyond a certain time. However, there is a quick workaround this: Suppose we already evaluated $\vec{x}(t)$ all the way till $t=T$, and would like to see beyond the radius of convergence. All we have to do is to reset the time to zero: we substitute ${\bf A}(T)$ and $\vec{x}(T)$ as initial conditions ${\bf A}_0$ and $\vec{x}_0$ and just use the same formula, now with a radius of convergence pushed forward in time. A principled way to determine the $T$ at which to do this replacement is given in  Appendix \ref{app:limits_of_theory}.

Using Eqn.\ref{x(t)} we can obtain the time to first extinction. We set $x_i(\tau)=0$, solve for $\tau$, and pick the $i$ that has the smallest $\tau$ value (for details, see Appendix \ref{app:relaxation_scheme_v2}),
\begin{equation}
    \tau_{m}=\min_i\frac{-[\vec{x}_0]_i}{[\vec{\epsilon}_1]_i + \sum_{j=2}^m[\vec{\epsilon}_j]_i \tau_{m-1}^{j-1}}.
    \label{eq:relax_v2_m_order}
\end{equation}
where $[\vec{\epsilon}_j]_i$ denotes the $i^\mathrm{th}$ component of the $j^\mathrm{th}$ $\vec{\epsilon}$ vector, and $\min_i()$ indicates to chose the value of $i$ that yields the smallest positive value of $()$. This is an iterated solution. We start by plugging in $\tau_0=0$ on the right hand side to get $\tau_1$. Then we plug in $\tau_1$ on the right hand side to get $\tau_2$, and so on, such that $\tau_m$ rapidly approaches to the true value of $\tau$ with increasing $m$.

Evaluating the first approximation $m=1$ for the time to extinction will turn out to be insightful,
\begin{equation}\tau_1=\min_i([\vec{x}_0]_i/[{\bf A_0^{-1}H_1}\vec{x}_0]_i).\label{tau1}\end{equation}
This result informs us of two risk factors for adaptation driven extinction: a small numerator and a large denominator. The first confirms what we might expect intuitively, that the species with the smallest abundances to begin with, are more likely to go extinct. The second risk factor is less intuitive. Species $i$ with large $[\vec{\epsilon}_1]_i=[{\bf A_0^{-1}H_1}\vec{x}_0]_i$ values are also more likely to go extinct. We tested the convergence of this scheme in Appendix \ref{app:relaxation_scheme_v2} which shows rapid convergence to the true extinction time when an extinction event occurs within the domain of convergence for Eqn.\ref{x(t)}.

Eqn.\ref{tau1} also suggests that larger communities will lose species faster, since the minimum of a larger list of numbers will be smaller. We confirm this trend in our numerical simulations (Fig.\ref{fig:CLG_fig}, first row).

We should emphasize that Eqn.\ref{x(t)} and Eqn.\ref{eq:relax_v2_m_order} (and the approximate Eqn.\ref{tau1}) are very general: As long as we are provided with ${\bf H}$'s we can track $\vec{x}(t)$ and determine extinction times regardless of the community structure, and regardless of the evolutionary mechanism that gives rise to the ${\bf H}$'s. In fact, the next section will be devoted to obtaining ${\bf H}$ matrices starting with the canonical equation of adaptive dynamics, which will be simply plugged into the above formulas. However, before doing so, let us illustrate a more practical use of the results we obtained so far.

Suppose that ${\bf A}$ is empirically measured at two times and then a line is fit, ${\bf A}={\bf A_0}+{\bf H_1}t$. In this special case, Eqn.\ref{x(t)} reduces to a geometric sum and can be evaluated in closed form,
\begin{equation}\vec{x}(t)=\vec{x}_0+{\bf U_1}t({\bf I}-{\bf U_1}t)^{-1}\vec{x}_0.\label{exactxwithU1}\end{equation}
where ${\bf U_1}=-{\bf A_0^{-1}}{\bf H_1}$. This formula is exact, and luckily, even has an infinite radius of convergence (Appendix \ref{app:constant_exact}). We compare it against numerical solutions and stochastic simulations and find excellent agreement (Fig.\ref{fig:const_xt}).

To obtain the time to first extinction $\tau$, we again set the left side of Eqn.$\ref{exactxwithU1}$ to zero, and identify which component $i$ crosses zero first (see Appendix \ref{app:relaxation_scheme_constant_exact})
\begin{equation}
    \tau_{m} = \min_i  \frac{-[\vec{x}_0]_i}{[{\bf U_1}({\bf I}-{\bf U_1}\tau_{m-1})^{-1}\vec{x}_0]_i}\label{tauconst}
\end{equation}
where again, we start the iteration with $\tau_0=0$ in the denominator of the right side, and $\min_i()$ indicates that the smallest positive value should be selected.

The first iteration for this special case gives us a $\tau_1$ that is identical to Eqn.\ref{tau1}, which of course is expected, since the general case depends only on ${\bf H_1}$ to begin with. 

Comparing Eqn.\ref{tauconst} to numerical simulations of an $N=100$ species community (see Numerical Methods) we find high accuracy and rapid convergence. The simple expression $\tau_1=\min_i(-[\vec{x}_0]_i/[{\bf U_1}\vec{x}_0]_i)$ gives us a reasonably good estimation to begin with, off only by $13\%$. The next iteration $\tau_2=\min_i(-[\vec{x}_0]_i/[{\bf U_1}({\bf I}-{\bf U_1}\tau_1)^{-1}\vec{x}_0]_i)$ brings this error down to $3\%$.

To better understand the qualitative behavior of Eqn.\ref{exactxwithU1} we write $\vec{x}_0$ as a linear combination of the eigenvectors of ${\bf U_1}$ (see Appendix \ref{app:relaxation_scheme_constant_exact})
\begin{equation}
\vec{x}(t) = \vec{x}_0+\sum_{j,k=1}^N \frac{({\bf V^{-1}})_{jk}[\vec{x}_0]_{k}\lambda_j t}{1-\lambda_j t} \vec{v}_j
\label{eq:const_eigen}
\end{equation}
where $\lambda_j$ and $\vec{v}_j$ are the $j^\mathrm{th}$ eigenvalue and eigenvector of ${\bf U_1}=-{\bf A_0^{-1}H_1}$; and ${\bf V}$ is a matrix formed by writing the $j^\mathrm{th}$ eigenvector $\vec{v}_j$ into the $j^\mathrm{th}$ column. We observe in Eqn.\ref{eq:const_eigen}, that the denominator approaches zero, as time approaches one over the largest eigenvalue. This means that the abundances of many species will either boom or crash simultaneously depending on the sign of the corresponding component of $\vec{v}_j$. 
Interestingly, we observed a similar singularity in a previous work that explored ways to edit community composition \cite{nguyen2020theoretical}. 

The first-order formula for extinction time $\tau_1$ is reasonably accurate, even though it can identify who goes extinct only $54\%$ of the time (although still better than random guessing, $1\%$). This is because many species start crashing around the same time. Fortunately, the second iteration $m=2$ increases the success rate to $93\%$. Since the first-order extinction time only takes into account the initial downward slope $\vec{\epsilon}_1$, this observation suggests that about half the time, the species initially in rapid decline are not the ones that goes extinct first. Half of the time, second-order effects (i.e., the evolution of neighbors of neighbors in the interaction network) will accelerate an otherwise slow linear decline, a scenario we mentioned in the introduction. 

Incidentally, if we are allowed to make two best guesses instead of one, extinctions can be predicted $71\%$ $(m=1)$ and $99\%$ $(m=2)$ of the time. 


Conveniently, our formulas for extinction time Eqn.\ref{eq:relax_v2_m_order} and Eqn.\ref{tauconst} need not be used only for the soonest extinction. To obtain the times for latter extinctions one simply removes the row and column from the interaction matrix corresponding to the species that went extinct, and substitute this reduced interaction matrix for ${\bf A_0}$ in Eqn.\ref{eq:relax_v2_m_order} and Eqn.\ref{tauconst}. 
This can be repeated to get a series of extinction times. Then these extinction times can be put together to obtain the number of extant species as a function of time $N(t)$ (Fig.\ref{fig:MC_Nt}).

In Fig.\ref{fig:const_xt}, \ref{fig:MC_Nt}, and first column of Fig.\ref{fig:CLG_fig}, we plot Eqn.\ref{exactxwithU1}, Eqn.\ref{tauconst} and the repeated use of the latter to get $N(t)$. We compare these results with mean-field simulations and stochastic simulations and find excellent agreement.

We should caution that the iterated use of Eqn.\ref{eq:relax_v2_m_order} and Eqn.\ref{tauconst} to obtain $N(t)$ neglects the possibility that extinct species can return back later. In other words, we remove species from the interaction matrix permanently, thereby neglecting the (seemingly rare) scenario where a community becomes uninhabitable for a species, after some time becomes habitable again, and the originally extinct species migrates or mutates back and successfully fixes.

\subsection{Canonical adaptive dynamics}

In the previous section we obtained how equilibrium abundances change for any community, given ${\bf H_1,H_2},\ldots$ In this section we will obtain these $\bf{H}$ matrices assuming the canonical equation of adaptive dynamics, which relates interaction values with the evolution of traits they depend on.

\begin{figure}
\centering
\includegraphics[width=\linewidth]{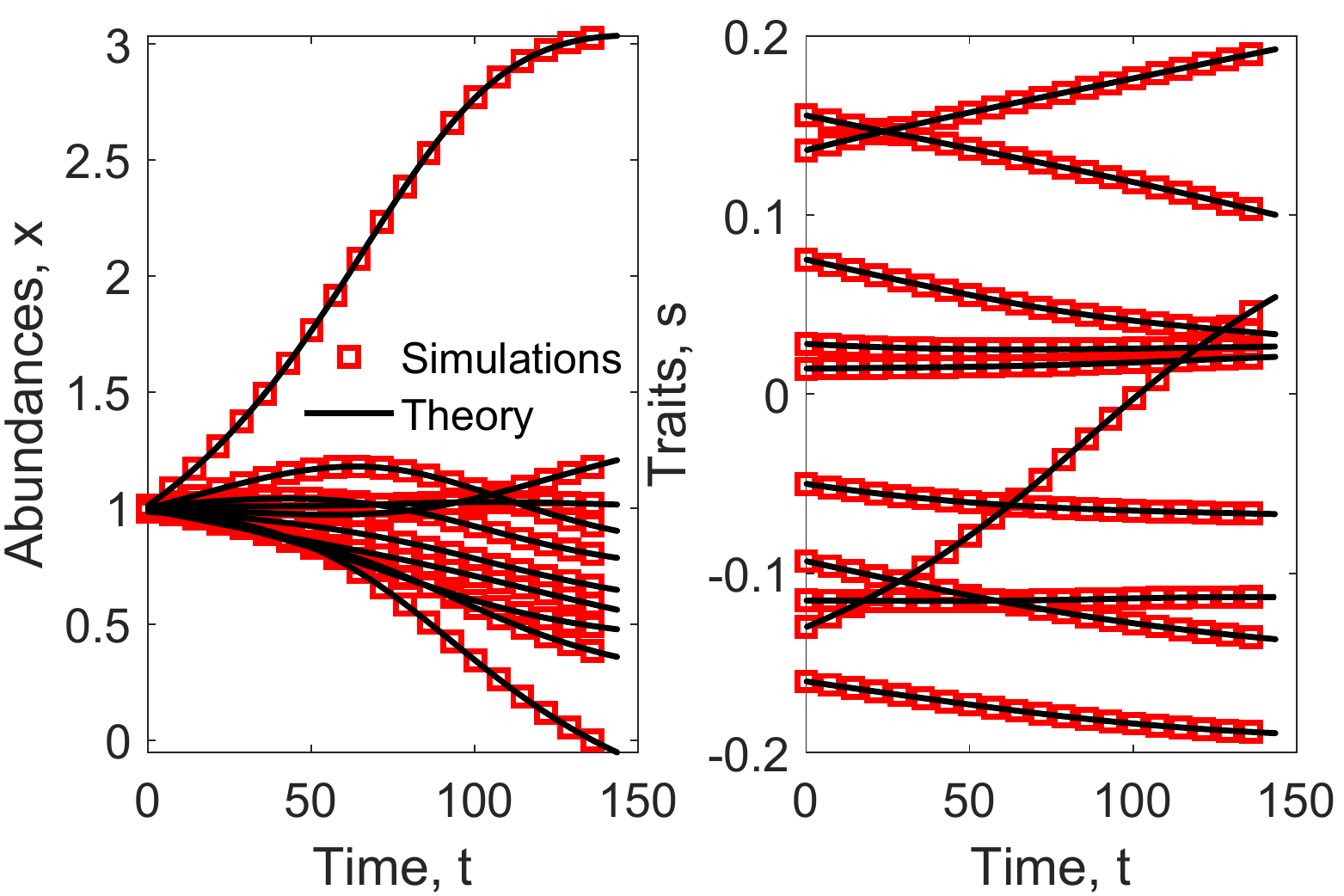}
\caption{{\bf Equilibrium abundances $\vec{x}(t)$ in an arms race adaptive dynamics community}. Our analytic formula (Eqn.\ref{x(t)}) correctly tracks each species abundance over time, and can also be used to predict the traits over time. The $N=50$ community was generated using the parameters presented in Numerical Methods using up to $O(t^4)$ for Eqn.\ref{x(t)} and \ref{eq:s(t)}. For visual clarity we only show the first $10$ species (unsorted). }
\label{fig:Linear_xt}
\end{figure}


Specifically, we assume that $A_{ij}(s_{11},s_{21},\ldots)$ depends on some traits $\vec{s}=\{s_{11}, s_{21},\ldots\}$. If at present, $t=0$ these traits are $\vec{s}_0$, for small changes in trait values we can Taylor expand
\begin{equation}
A_{ij}(\vec{s})\approx A_{ij}(\vec{s}_0) + \sum_{k\alpha}\frac{\partial A_{ij}}{\partial s_{k\alpha}}\bigg|_{\vec{s}_0}\delta s_{k\alpha} \label{taylor}
\end{equation}
where $\delta s_{k\alpha}=(ds_{k\alpha}/dt)\delta t$ is the amount that the $\alpha^\mathrm{th}$ trait of the $k^\mathrm{th}$ species can change within some time interval $\delta t$. The time interval for which this approximation is valid depends on the functional form of $A_{ij}(\vec{s})$. 

Then we write the chain rule
\begin{equation}
\frac{dA_{ij}}{dt} = \sum_{k\alpha} \frac{\partial A_{ij}}{\partial s_{k\alpha}}\frac{ds_{k\alpha}}{dt},
\end{equation}
plug in Eqn.\ref{canonical} for $ds_{k\alpha}/dt$, with $f_k=r_k+\sum_lA_{kl}x_l$, expand $\vec{x}(t) = \vec{\epsilon}_1 + \vec{\epsilon}_1 t + \vec{\epsilon}_2 t^2 + ...$ and $
    {\bf A}(t) = {\bf H_0} + {\bf H_1} t + {\bf H_2} t^2 + ...
$ and match polynomial coefficients to get (Appendix \ref{app:dAdt_and_H})
\begin{equation}
    [{\bf H_k}]_{ij} = \sum_{n q\alpha}\frac{\mu_{n\alpha}\sigma_{n\alpha}^2}{2 k \lambda} C_{ijn\alpha} \sum_{s=0}^{k-1} [\vec{\epsilon}_s]_{n} C_{nqn\alpha}[\vec{\epsilon}_{k-s-1}]_{q}\label{adresult}
\end{equation}
where $C_{ijk\alpha}=\partial A_{ij}/\partial s_{k\alpha}$ depends on the specifics of the biology (i.e., how interactions depend on traits). The indices $n$ and $q$ run over all the species, whereas $\alpha$ runs over all the traits. Note that in this formula, each order of ${\bf H}$ depends on previous orders of $\vec{\epsilon}$'s, which themselves depend on previous orders of ${\bf H}$, via Eqn.\ref{implicit}. Thus, Eqn.\ref{adresult} can be evaluated order by order, starting with ${\bf H_0}={\bf A_0}$ and $\vec{\epsilon}_0=\vec{x}_0$.

This is our central result. With ${\bf H}$'s in hand, one evaluates Eqn.\ref{x(t)} and Eqn.\ref{eq:relax_v2_m_order} to get abundances $\vec{x}(t)$, extinction times $\tau$, and number of extant species $N(t)$. Furthermore, now that we know $\vec{x}(t)$, we can plug it in Eqn.\ref{canonical} and obtain the evolution of traits, 
\begin{align}
    s_{k\beta}(t) = s_{k\beta}(0) + \frac{\mu_{k\beta}\sigma_{k\beta}^2}{2\lambda}\sum_{\alpha, q}\bigg[[\vec{x}_0]_k C_{kqk\alpha}[\vec{x}_0]_q t\nonumber\\
    +\big([\vec{x}_0]_k C_{kqk\alpha}[\vec{\epsilon}_1]_q+[\vec{\epsilon}_1]_k C_{kqk\alpha}[\vec{x}_0]_q\big) \frac{t^2}{2}+...\bigg]
    \label{eq:s(t)}
\end{align}
where $s_{k\beta}$ is the $\beta$-th trait for species $k$. In Fig.\ref{fig:Linear_xt} we compare numerical simulations (cf. Applications and Numerical Methods sections) to these analytical formulae and find excellent agreement.

\subsection{Applications}
We now use our central result, Eqn.\ref{adresult} (as plugged into Eqn.\ref{x(t)} and \ref{eq:relax_v2_m_order}) to solve some classical models gathered from the AD literature. All we need to do is to extract $C_{ijkl}$ from a specific model, and plug it into Eqn.\ref{adresult}. We will illustrate how to do this for two models. In both models, species happen to have only have one evolving trait, so $\alpha$ can only be 1 in $C_{ijk\alpha}$, and $s_{i\alpha}$. For brevity, we will omit writing 1 each time, $C_{ijk1}\rightarrow C_{ijk}$ and $s_{i1}\rightarrow s_i$.

{\bf 1. Escalating Traits (Linear A(s))}. We start with the model in \cite{schaffer1978homage,rosenzweig1978homage,rosenzweig1987red}, which assumes an arms race scenario between every predator-prey pair. Over evolutionary time scales, each species develops their trait to counter the progress of their opponent. This has been empirically shown by Brodie and Brodie through the correlation between the toxin-resistance of garter snakes and toxicity of newts in various geographical areas \cite{brodie1990tetrodotoxin,brodie1991evolutionary,brodie1999costs,brodie1999predator}. Their work suggests snakes have greater toxin resistance in order to consume newts with higher toxicity.

\begin{figure}
\centering
\includegraphics[width=\linewidth]{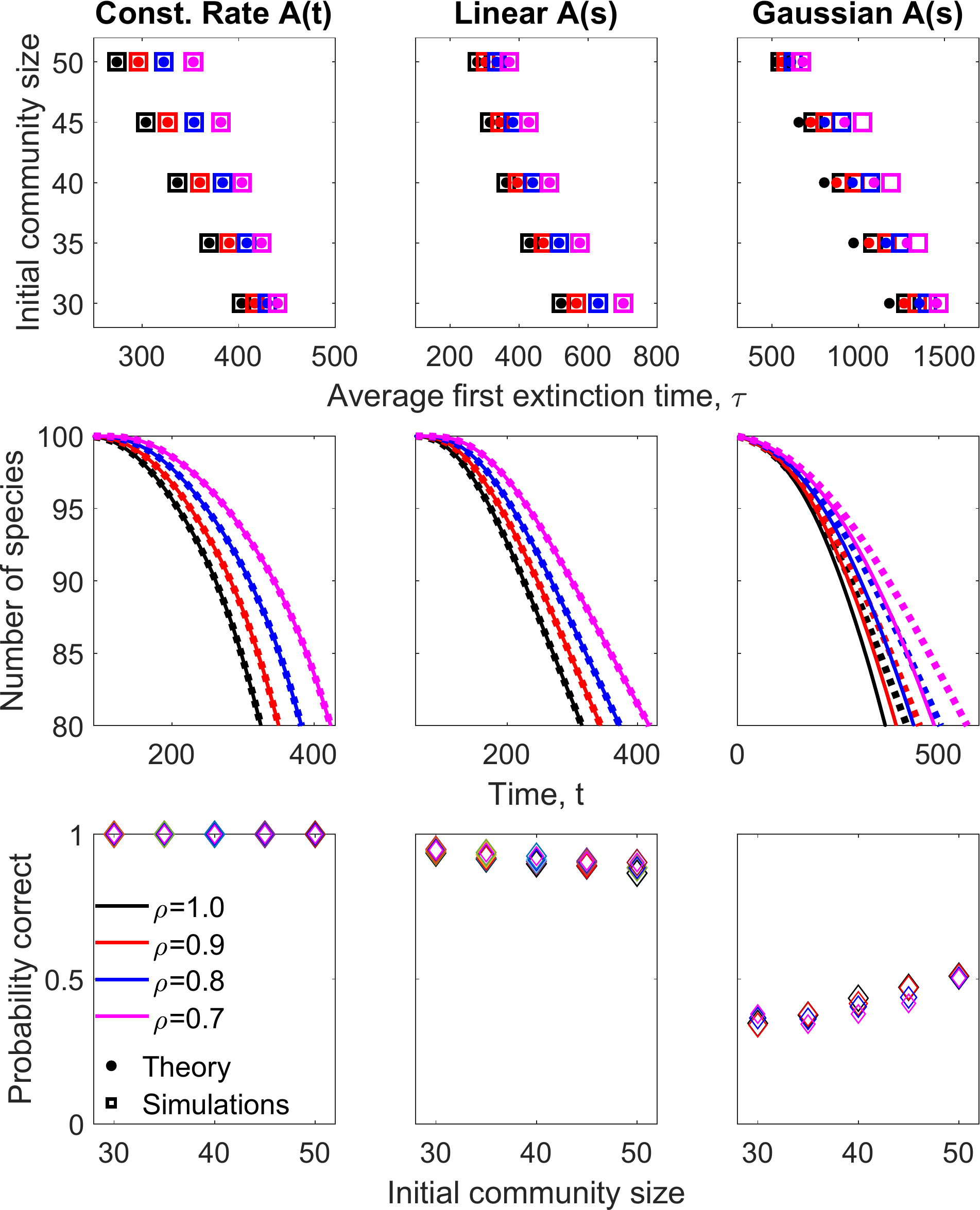}
\caption{{\bf Adaptive Extinction.} We gathered classical AD models from the literature \cite{schaffer1978homage,rosenzweig1978homage,rosenzweig1987red,marrow1992coevolution,marrow1996evolutionary}, and compared their specific numerical solutions to our general formulae. To do so we sampled $5000$ communities starting with varying initial number of species and interaction densities $\rho=(0.7,0.8,0.9,1.0)$ as (black, red, blue, magenta), respectively. The results from adaptive dynamics (AD) are shown as squares and iterating the theory (middle and right columns) gives the results shown as solid/filled circles and the constant rate ${\bf A}(t)$ case (left column) uses the exact solution shown in Eqn.\ref{eq:const_eigen}. Each simulation started with a community of size between $N=30$ and $N=100$ for interaction densities $\rho=[0.7,0.8,0.9,1.0]$.
{\bf Top:} average first extinction time.
{\bf Middle:} number of surviving species over time.
{\bf Bottom:} probability of correctly predicting the first extinct species.}
\label{fig:CLG_fig}
\end{figure}

The interaction between $i$ and $j$ depends on the difference between their two traits $A_{ij}(s_i,s_j) = (s_i-s_j)M_{ij}$ and $A_{ji}(s_i,s_j) = (s_i-s_j)M_{ji}$. Note that reciprocal interactions use the same trait difference but can have different signs (due to ${\bf M}$). The first term scales the strength of the interaction with the difference in traits, and allows for the scenario where this difference changes in sign. For example, if we were modeling two types of carnivorous fish, if a prey species evolves to be larger than its predator then the role of predator and prey can be flipped.

Next, we evaluate $C_{ijk}=\partial A_{ij}/\partial s_k$. For all $i<j$,
\begin{align*}
    C_{ijk} = M_{ij}(\delta_{ik}\!-\!\delta_{jk}), C_{jik} = M_{ji}(\delta_{ik}\!-\!\delta_{jk}),\mbox{and}\
    C_{iik} = 0
\end{align*}
where $\delta_{ik}$ is the Kronecker delta. We plug these into Eqn.\ref{adresult} and compare the analytical model against numerical simulations of the model, and find excellent agreement (Fig.\ref{fig:CLG_fig}, middle column). 

{\bf 2. Bounded traits (Gaussian A(s)).}
Next, we consider the model in \cite{marrow1992coevolution,marrow1996evolutionary} where interactions are bounded, such that $A_{ij}(\vec{s})$ has a Gaussian peak. Specifically,
\begin{align}
    &A_{ij}(s_i,s_j) = M_{ij}\exp\left[\!-\!\beta_{ij}^2\!-\!\gamma_{ij}^2\!+\!2 [{\bf P_5}]_{ij} \beta_{ij}\gamma_{ij}\right],\\
    &\beta_{ij}=(s_i\!-\![{\bf P_1}]_{ij})/[{\bf P_2}]_{ij},\mbox{and}\
    \gamma_{ij}=(s_j\!-\![{\bf P_3}]_{ij})/[{\bf P_4}]_{ij}\nonumber
\end{align}
where all model parameters are contained inside the matrices ${\bf P_1},{\bf P_2},\dots,{\bf P_5}$. Similar to the linear ${\bf A}(\vec{s})$ model, we use ${\bf M}$ to incorporate the signs, trade-off, and density of interactions. In this model the interactions cannot change signs so one can incorporate a more realistic predator-prey type of interactions by restricting the positive interactions to be weaker than the negative interactions (cf. Numerical Methods). 

Next, we evaluate $C_{ijk}=\partial A_{ij}/\partial s_k$. For all $i<j$
\begin{align}
    C_{ijk} = &\delta_{ik} \frac{2A_{ij}}{[{\bf P_2}]_{ij}}\left([{\bf P_5}]_{ij} \frac{s_j-[{\bf P_3}]_{ij}}{[{\bf P_4}]_{ij}}\!-\!\frac{s_i-[{\bf P_1}]_{ij}}{[{\bf P_2}]_{ij}}\right) \nonumber\\
    + &\delta_{jk} \frac{2A_{ij}}{[{\bf P_4}]_{ij}}\left([{\bf P_5}]_{ij} \frac{s_i-[{\bf P_1}]_{ij}}{[{\bf P_2}]_{ij}}\!-\!\frac{s_j-[{\bf P_3}]_{ij}}{[{\bf P_4}]_{ij}}\right),\nonumber\\
    C_{jik} = &\delta_{jk} \frac{2A_{ji}}{[{\bf P_4}]_{ji}}\left([{\bf P_5}]_{ji} \frac{s_i-[{\bf P_1}]_{ji}}{[{\bf P_2}]_{ji}}\!-\!\frac{s_j-[{\bf P_3}]_{ji}}{[{\bf P_4}]_{ji}}\right) \nonumber\\
    + &\delta_{ik} \frac{2A_{ji}}{[{\bf P_2}]_{ji}}\left([{\bf P_5}]_{ji} \frac{s_j-[{\bf P_3}]_{ji}}{[{\bf P_4}]_{ji}}\!-\!\frac{s_i-[{\bf P_1}]_{ji}}{[{\bf P_2}]_{ji}}\right),\nonumber\\
    \mbox{and}\nonumber\\
    C_{iik} = &0\nonumber.
\end{align}

As each species changes their trait monomorphically we update the values of ${\bf A}$ according to each interaction model and then determine the population abundances using the Eqn.\ref{eq:fixedpoint}. These numerical simulations are compared against our analytical formula Eqn.\ref{adresult} in the third column of Fig.\ref{fig:Linear_xt}. 

As expected, analytical theory agrees with simulations during some initial time, and then gradually departs (Fig.\ref{fig:CLG_fig}, third column). This is because of our linearity assumption Eqn.\ref{taylor}. The Gaussian ${\bf A}(\vec{s})$ can be approximated by a line only for small trait changes. Of course, one could remedy this by occasionally correcting the slope of ${\bf A}(\vec{s})$. However, even without doing so, our formula retains its accuracy over a timescale that spans many extinctions (Fig.\ref{fig:CLG_fig}, third row, second column). Note however, that we can accurately pinpoint which species will go extinct only about half the time (Fig.\ref{fig:CLG_fig}, third row, third column).

\section{Numerical methods}
We have applied our analytical formulae to solve some models gathered from the literature, which so far had only specific numerical solutions. In this section we outline the initial conditions and parameter values we chose for reproducing these numerical solutions, for the purpose of comparing them against our analytical formulae.

All initial communities were generated using normal distribution for the population abundances and trait parameters for varying community sizes and interaction types. Across the three sampled models we use the same average population abundance $\langle x\rangle=1$ and standard deviation $\langle \sigma_x \rangle\ =0.1$. For the two models which simulates AD traits we initialized the traits $\langle s \rangle=0$ with standard deviation $\langle\sigma_s\rangle=0.1$. The classification matrix ${\bf M}$ randomly assigns predator-prey interactions for each pair $(i,j)$. Additionally, we modify the interaction density, $0\leq\rho\leq 1$, which is the probability that we do not assign $M_{ij}=M_{ji}=0$ to a matrix element.

The initial communities for the constant rate ${\bf A}(t)$ uses the interactions generated by the linear ${\bf A}(\vec{s})$ but does not simulate the traits. Instead we randomly sampled the values of ${\bf H}$ from a normal distribution $H_{ij} = \mu\times N[1,1]$ for $\mu=10^{-4}$ and kept self-competition and initially non-existent interactions from changing. The linear ${\bf A}(\vec{s})$ model uses $\mu_{k}=10^{-3}$, the Gaussian ${\bf A}(\vec{s})$ model uses $\mu_{k}=10^{-2}$, and both of these model use $\sigma_{k}=10^{-3}$ and $\lambda_k=10^{-3}$ for all species $k$. The stochastic trajectories were generated for Fig.\ref{fig:const_xt} and \ref{fig:MC_Nt} by assigning each off-diagonal element a probability of change, $0<\zeta\leq 1$ per unit time. In order to keep the average change over time the same the rate of change was modified to $H_{ij}\rightarrow H_{ij}/\zeta$ so as $\zeta$ decreased from unity each element experiences rare but stronger changes than the mean-field. 

The Gaussian ${\bf A}(\vec{s})$ model requires additional parameters for its peaks (${\bf P_1}$, ${\bf P_3}$) and widths (${\bf P_2}$, ${\bf P_4}$). These parameters are symmetric for each interaction pair ($P_{ij}=P_{ji}$) with an average peak magnitude $\langle |{\bf P_1}|\rangle=\langle |{\bf P_3}|\rangle=5$ and widths $\langle {\bf P_3}\rangle=\langle {\bf P_4}\rangle=10$. The peak values have a standard deviation of $0.1$ and are randomly chosen as either positive or negative with equal probability. The width values have unit standard deviation and are strictly positive. The coupling coefficients (${\bf P_5}$) for each interaction pair is uniformly random between zero and unity while keeping it symmetric. The sign of each Gaussian interaction cannot change as the community evolves its trait so positive interaction values $A_{ij}>0$ are given an inefficient mass-transfer penalty of $10\%$ in comparison to the negative interaction values ($A_{ij}=-0.1 A_{ji}$). The overall magnitude of the Gaussian interactions are scaled up by a factor of $50$ in order to accelerate changes.

Similar to \cite{marrow1996evolutionary} we assume that the intrinsic birth/death vector does not change with mutations for both models. However, we also assume that the intraspecific interaction for each species is kept constant. In contrast, \cite{marrow1996evolutionary} allows the intraspecific interaction of the prey to change with its trait, and there is no self-limiting interaction for the predator. To keep our model simple, the self-interaction for all species is kept at $-10$.

In our models we assume that a random walk in interaction space replicates a random walk in trait space of AD in a low mutation rate and high selection limit. Comparing the stochastic effect of this random walk to our deterministic mean-field model (Fig.\ref{fig:const_xt} and \ref{fig:MC_Nt}) shows that our deterministic (mean-field) approximations agrees with the average stochastic behavior. Therefore, we can average over the stochastic effects of randomly walking in interaction space by just using our mean-field theory.



\section{Conclusion}
We obtained general analytical formulas for species abundances (Eqn.\ref{x(t)} and \ref{exactxwithU1}), evolution of interaction mediating traits (Eqn.\ref{eq:s(t)}), and a series of extinction times (Eqn.\ref{eq:relax_v2_m_order} and \ref{tauconst}) for complex communities involved in multi-party arms races. We then showed how to evaluate these formulas when the species interactions are governed by the standard adaptive dynamics equations (Eqn.\ref{adresult}), applied these analytical formulas to solve AD models gathered from the literature \cite{schaffer1978homage,rosenzweig1978homage,rosenzweig1987red,marrow1992coevolution,marrow1996evolutionary}, and compared our analytical solutions to specific numerical solutions of these models. 

Having analytical formulas, as opposed to numerical solutions, show us the effect of every single parameter on the abundances and series of extinctions. For example, we find that a crucial quantity governing the extinction is $[{\bf A_0^{-1} H_1}\vec{x}_0]_i$ (for species $i$), which is the most prominent factor in almost all of our results. We can also see the precise dependence of relative initial populations on extinction time, as well as the influence of community size. Another interesting observation is, if all initial populations are modified by a constant factor, then Eqn.\ref{eq:relax_v2_m_order} shows us that $\vec{x}(t)$ scales by the same amount, but the extinction times (which after all, are obtained by setting the left hand zero), remain invariant. In other words, communities with small abundances exhibit similar adaptive extinction times with those with large abundances.

In closing, we mention two important problems that still remain open. In the present work we analytically obtained a string of extinction times, which we stitched together to obtain the number of extant species $N(t)$ as a function of time. However, we were not able to obtain a closed form formula for $N(t)$ that would make apparent the conditions for which a community would approach to a fixed number of stably coexisting species (in the red queen sense), versus vanish entirely. For communities that do approach a stable number of coexisting coevolving species, we would like to know what this number is, as a function of initial community size and interaction parameters. 

Such a result might also provide some insight into May's paradox \cite{may1972will}, which, as many authors before us have pointed out, can be resolved by structured (as opposed to random) interaction matrices. 

The evolutionary process that structure interaction matrices is adaptive dynamics. Thus, having a closed form formula for $lim_{t\to\infty}N(t)$ would allow us to catalogue the interaction structures that allow non-vanishing communities, and those that lead to maximal (possibly, even divergent) number of species. It would also allow us to see what, if any, is the upper bound to the number of species that can coexist while also coevolving.

The second open problem is the following. Presently, our adaptive dynamics solutions accept as inputs, the slope of the function ${\bf A}(\vec{s})$. In other words, one must empirically measure how interactions change with respect to some traits, and fit lines. It seems to us that it should be feasible to generalize the calculations here to any arbitrary function of ${\bf A}(\vec{s})$, which then would allow evolutionary predictions that hold for a much longer time.

The main predictive challenge in testing the feedback between evolutionary and population dynamical timescales lies in determining the mapping from traits to interactions. It is currently not feasible to fully determine $N\times N$ functions, ${\bf A}(\vec{s})$, empirically. However, it \emph{may} be feasible to measure the local rates of change in interactions $d{\bf A}/dt$. As such, our implicit AD approach, Eqn.\ref{x(t)} could allow for an easier (albeit less mechanistic) integration of observation and theory. When our analytical model is used with the local rates of change in the slopes, we can accurately track the abundance levels over long periods of time, albeit not indefinitely. However, coupling analytical theory with periodic measurements of the local values of $d{\bf A}/dt$ one can avoid mapping traits to interactions and still retain high accuracy in predicting abundance trajectories and series of extinctions.

\section{Conflict of Interest}
The authors declares no conflict of interests.
\section{Data Availability Statement}
The MATLAB code used to generate the data for our paper can be found in the supporting information at: https://doi.org/10.1111/jeb.13796

\bibliography{bibliography.bib}
\bibliographystyle{unsrt}

\appendix
\section{Implicit adaptive dynamics - Abundances}
\label{app:general_approximation}
For a time-dependent interaction matrix, ${\bf A}(t)$, we wish to find a time-dependent form for the equilibrium population abundances, $\vec{x}(t)$, through the fixed-point constraint ${\bf A}\vec{x}=-\vec{r}$ given the intrinsic birth/death rates $\vec{r}$. Then by expanding the time-dependencies as a polynomial with respect to time, $t$, we can then write
\begin{equation*}
    \vec{x}(t) = \vec{\epsilon}_0 + \vec{\epsilon}_1 t + \vec{\epsilon}_2 t^2 + \dots
\end{equation*}
and assume that the form of ${\bf A}(t)$ is also given as
\begin{equation}
    {\bf A}(t) = {\bf H_0} + {\bf H_1} t + {\bf H_2} t^2 + \dots
    \label{eq:A_Ht}
\end{equation}
where all coefficients (${\bf H},\vec{\epsilon}$) are constant with respect to time. This then gives
\begin{equation*}
    \left({\bf H_0} + {\bf H_1} t + {\bf H_2} t^2 + \dots\right)\left(\vec{\epsilon}_0 + \vec{\epsilon}_1 t + \vec{\epsilon}_2 t^2 + \dots\right)=-\vec{r}
\end{equation*}
equating each power of $t$ gives
\begin{align*}
    {\bf H_0}\vec{\epsilon}_0 &= -\vec{r}\\
    {\bf H_0}\vec{\epsilon}_1+{\bf H_1}\vec{\epsilon}_0 &= 0\\
    {\bf H_0}\vec{\epsilon}_2+{\bf H_1}\vec{\epsilon}_1+{\bf H_2}\vec{\epsilon}_0 &= 0\\
    \vdots\\
    \quad \sum_{k=0}^m {\bf H}_{k}\vec{\epsilon}_{m-k} &= 0
\end{align*}
where $m>0$. The zeroth-order gives our starting point by definition ${\bf H}_0 = {\bf A}(t=0)$ and $\vec{\epsilon}_0 = \vec{x}(t=0)$. We then sequentially calculate the higher-order coefficients for the population abundances as
\begin{align*}
    \vec{\epsilon}_1 &= -{\bf H_0^{-1}}{\bf H_1}\vec{\epsilon}_0\\
    \vec{\epsilon}_2 &= -{\bf H_0^{-1}}\big({\bf H_1}\vec{\epsilon}_1+{\bf H_2}\vec{\epsilon}_0\big)\\
    \vdots\\
    \vec{\epsilon}_m &= -{\bf H_0^{-1}}\sum_{k=1}^{m} {\bf H}_{k}\vec{\epsilon}_{m-k}
\end{align*}
where each coefficient only depends on terms of a lower-order. This will also work for population dependent mutation rates as long as we can write ${\bf H_m}$ using terms up to order $m-1$.

For example, if the interactions are only linearly changing with time then all higher-orders are removed, ${\bf H_2} = {\bf H_3} = \dots = 0$, and the coefficients can be simplified to
\begin{align}
    \vec{\epsilon}_1 &= -{\bf H_0^{-1}}{\bf H_1}\vec{\epsilon}_0\nonumber\\
    \vec{\epsilon}_2 &= -{\bf H_0^{-1}}{\bf H_1}\vec{\epsilon}_1 = (-{\bf H_0^{-1}}{\bf H_1})^2\vec{\epsilon}_0\nonumber\\
    \vdots\nonumber\\
    \vec{\epsilon}_m &= (-{\bf H_0^{-1}} {\bf H_1})^m\vec{\epsilon}_{0}.\label{eq:eps_for_constant_H}
\end{align}
\section{Implicit adaptive dynamics - Time to extinction}
\label{app:relaxation_scheme_v2}

In this section we will develop a scheme to easily estimate the extinction time for each species as we iteratively increase the accuracy of the polynomial expansion of $\vec{x}(t)$. Given that ${\bf A}(t) = {\bf A_0}+{\bf H_1}t+{\bf H_2}t^2+\dots $ we can track the population abundances over time using the coefficients given by Eqn.\ref{eq:eps_for_constant_H}. This gives
\begin{equation*}
    \vec{x}(t) = \vec{x}_0+\vec{\epsilon}_1 t + \vec{\epsilon}_2 t^2 + \dots
\end{equation*}
which can be solved to find the extinction time for each species $x_i(\tau)=0$ and pick the earliest extinction time. However, this requires that the series be truncated at some order $t^m$ and the calculated extinction time will vary slightly as the order is changed. We wish to quickly estimate the extinction time using an $m$-order polynomial by iteratively calculating the extinction time $\{\tau_1,\tau_2,\dots,\tau_m\}$ as we increase the time order $m$. As we increase the order we should quickly converge to the true extinction time by relying on our previous estimates without needing to solve for the roots of a high-order polynomial. The basic idea is to convert the equations from $0=[\vec{x}]_i(\tau)$ to
\begin{equation}
    0=[\vec{x}_0]_i + [\vec{\epsilon}_1]_i\tau_m + [\vec{\epsilon}_2]_i\tau_m\tau_{m-1} + \dots + [\vec{\epsilon}_m]_i\tau_m\tau_{m-1}^{m-1}
    \label{eq:relaxation_v2_basis}
\end{equation}
by replacing $\tau^k = \tau_m\tau_{m-1}^{k-1}$ such that the highest-order extinction time $\tau_m$ appears in every power of time and can be easily solved for by factoring it out for each species $i$ to solve for the earliest positive extinction time. This approximation should then converge as $\tau_m\rightarrow\tau$.

For a community of $N$ species this system of equations has $N$ possible extinction times. There will be an extinction time for each species, however, there cannot be multiple extinction times for the community. Instead, the species with the earliest extinction event will die and the community structure will rearrange after its death. So as we are iterating this method we must pick for this specific extinction time. In other words, given the $N$ possible extinction times returned by the iteration we will pick the earliest \emph{positive} extinction time. 

For $m=1$ we have for the $i^\mathrm{th}$ species
\begin{equation*}
    0 = [\vec{x}_0]_i + [\vec{\epsilon}_1]_i \tau_{1i}
\end{equation*}
to get
\begin{equation}
    \tau_{1i} = \frac{-[\vec{x}_0]_i}{[\vec{\epsilon}_1]_i}.
    \label{eq:relax_v2_linear}
\end{equation}
where we discard any negative extinction times and pick the earliest positive time, $\tau_1=\min_i \tau_{1i}$, as our first-order approximation to the extinction time. Then for $m=2$
\begin{equation*}
    0 = [\vec{x}_0]_i + [\vec{\epsilon}_1]_i \tau_{2i} + [\vec{\epsilon}_2]_i \tau_{2i} \tau_1
\end{equation*}
to get
\begin{align}
    \tau_{2i} &= \frac{-[\vec{x}_0]_i}{[\vec{\epsilon}_1]_i + [\vec{\epsilon}_2]_i\tau_1}
    \label{eq:relax_v2_quadratic}
\end{align}
where we plug-in the result for $\tau_1$, and then again discard the negative extinction times and pick the earliest positive, $\tau_2=\min_i \tau_{2i}$. This procedure can be iterated $m$ times to get
\begin{equation}
    \tau_{mi}=\frac{-[\vec{x}_0]_i}{[\vec{\epsilon}_1]_i + \sum_{j=2}^m[\vec{\epsilon}_j]_i \tau_{m-1}^{j-1}}.
\end{equation}
and applying our positive minimization at each step.

Assuming that the true extinction time is $\tau$ then each iteration of this method has an error $\delta_m$ such that $\tau_m = \tau + \delta_m$. Assuming that the errors are small relative to the true extinction time, $\delta_m/\tau\ll1$ we can then estimate the error ratio $\delta_{m}/\delta_{m-1}$ by using the approximation
\begin{equation*}
    \tau_m^k=(\tau+\delta_m)^k\approx\tau^k(1+\frac{k\delta_m}{\tau}).
\end{equation*}
Then for $\tau_{m}$ and the extinction of the $i^\mathrm{th}$ species we have
\begin{align}
    \tau+\delta_{m} &= \frac{-[\vec{x}_0]_i}{[\vec{\epsilon}_1]_i+\sum_{j=2}^{m}[\vec{\epsilon}_j]_i(\tau+\delta_{m-1})^{j-1}} \nonumber\\
    &\approx \frac{-[\vec{x}_0]_i}{[\vec{\epsilon}_1]_i+\sum_{j=2}^{m}[\vec{\epsilon}_j]_i\left(\tau^{j-1}+(j-1)\delta_{m-1}\tau^{j-2}\right)}
    \label{eq:relaxation_v2_convergence_condition_1}
\end{align}
Now we assume that the expansion is sufficiently high-order to closely approximate the true extinction time such that
\begin{equation*}
    0\approx [\vec{x}_0]_i + [\vec{\epsilon}_1]_i \tau + [\vec{\epsilon}_2]_i \tau^2 + \dots+[\vec{\epsilon}_m]_i\tau^m
\end{equation*}
We can then multiply Eqn.\ref{eq:relaxation_v2_convergence_condition_1} on the top and bottom with $\tau$ to simplify to
\begin{equation*}
    \tau+\delta_m \approx \frac{-[\vec{x}_0]_i\tau}{-[\vec{x}_0]_i+\sum_{j=2}^m(j-1)[\vec{\epsilon}_j]_i\tau^{j-1}\delta_{m-1}}.
\end{equation*}
Now assuming the residue terms are small relative to the initial population sizes we can then again apply the binomial approximation to get
\begin{equation*}
    \tau+\delta_m \approx \tau\left(1+\frac{1}{[\vec{x}_0]_i}\sum_{j=2}^m(j-1) [\vec{\epsilon}_j]_i\tau^{j-1}\delta_{m-1}\right)
\end{equation*}
so the convergence ratio is given by
\begin{equation}
    \left|\frac{\delta_m}{\delta_{m-1}}\right|\approx\frac{1}{[\vec{x}_0]_i}\left|\sum_{j=2}^m(j-1)[\vec{\epsilon}_j]_i\tau^{j}\right|.
    \label{eq:relaxation_v2_convergence_final}
\end{equation}
This can be evaluated before hand for each species to determine the convergence of their extinction times. For our system the coefficients $\epsilon$ are very small relative to the initial population abundances which allows this method to rapidly converge.

\begin{figure}
    \centering
    \begin{subfigure}
        \centering
        \includegraphics[width=0.46\linewidth]{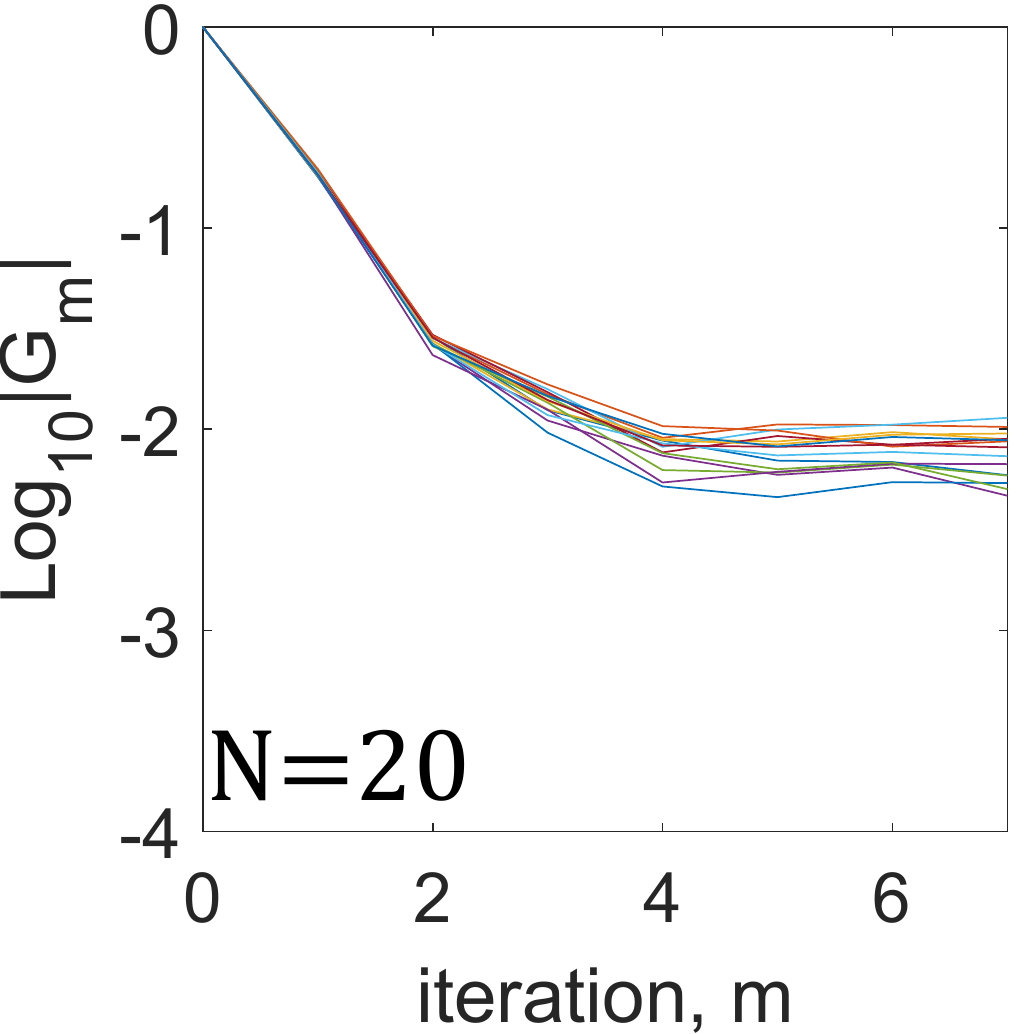}
    \end{subfigure}
    \begin{subfigure}
        \centering
        \includegraphics[width=0.46\linewidth]{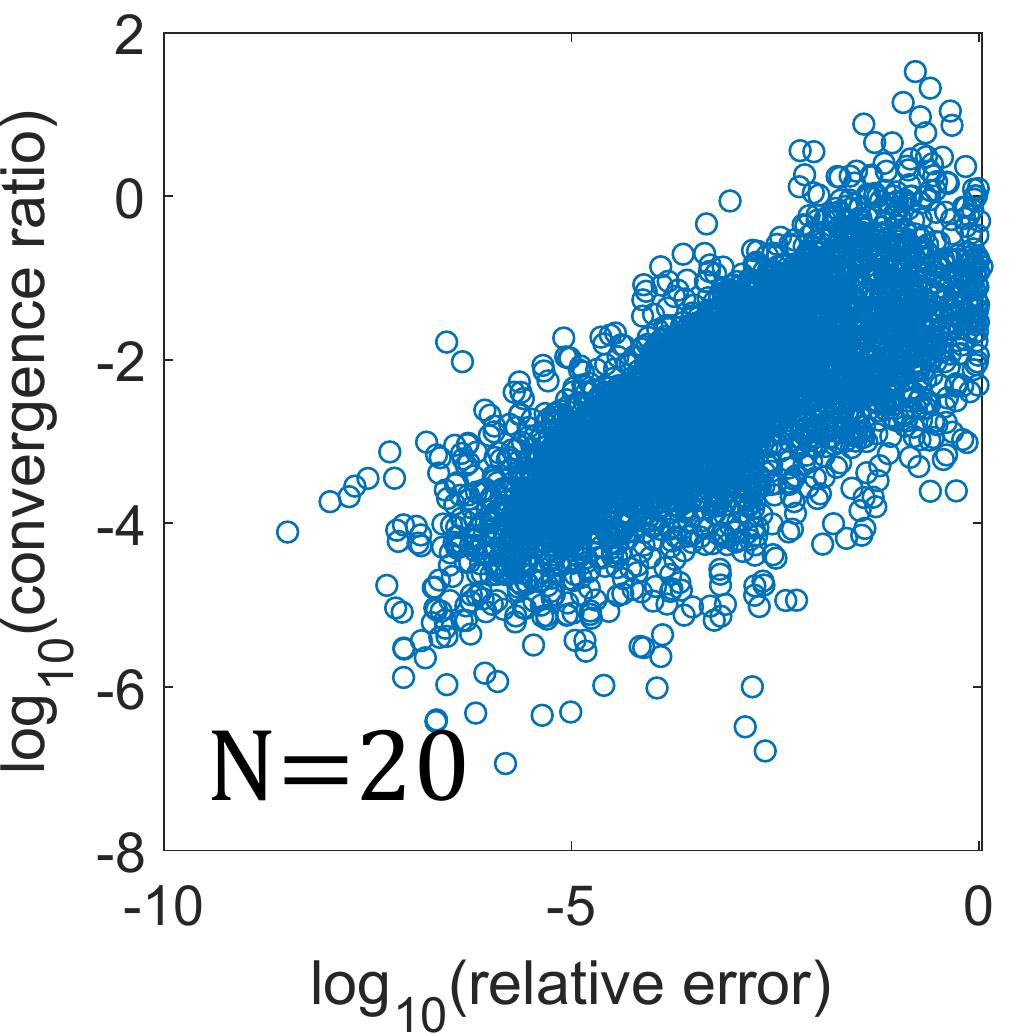}
    \end{subfigure}\begin{subfigure}
        \centering
        \includegraphics[width=0.46\linewidth]{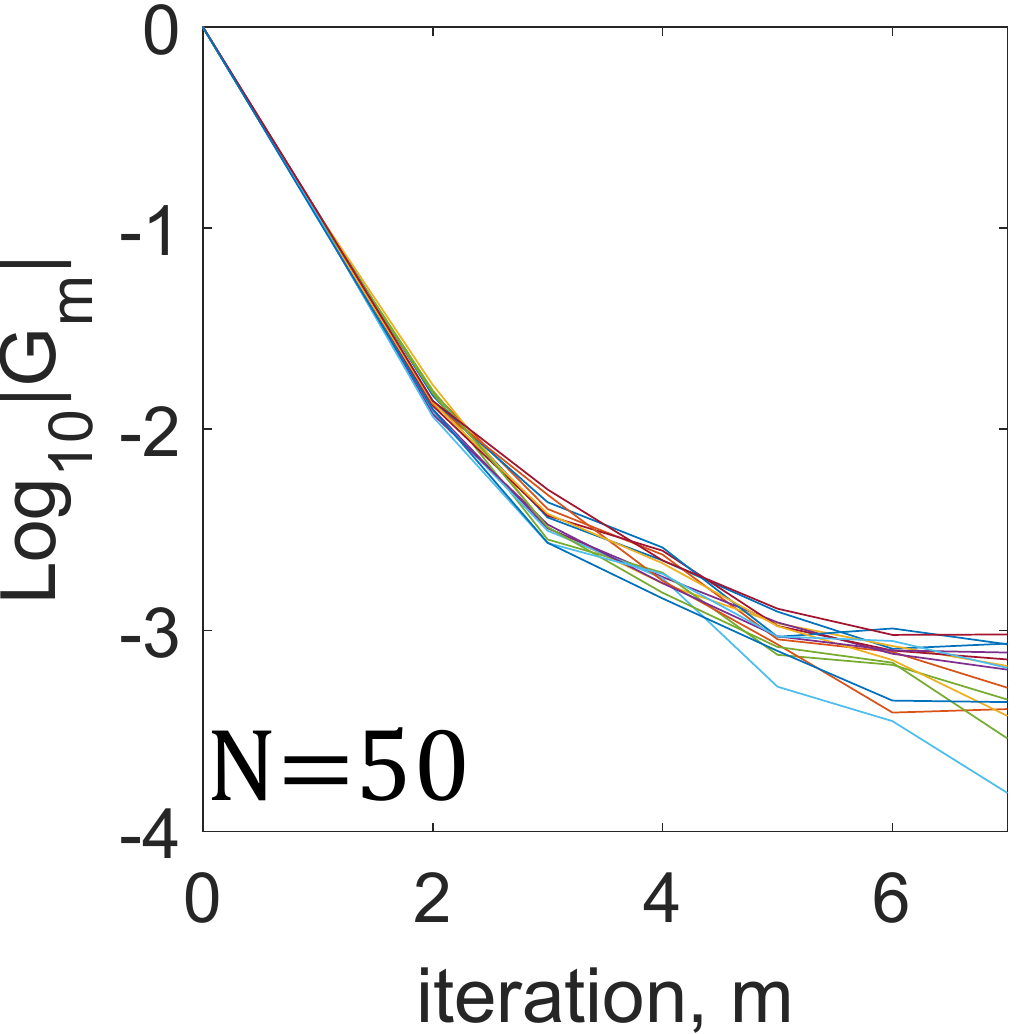}
    \end{subfigure}\begin{subfigure}
        \centering
        \includegraphics[width=0.46\linewidth]{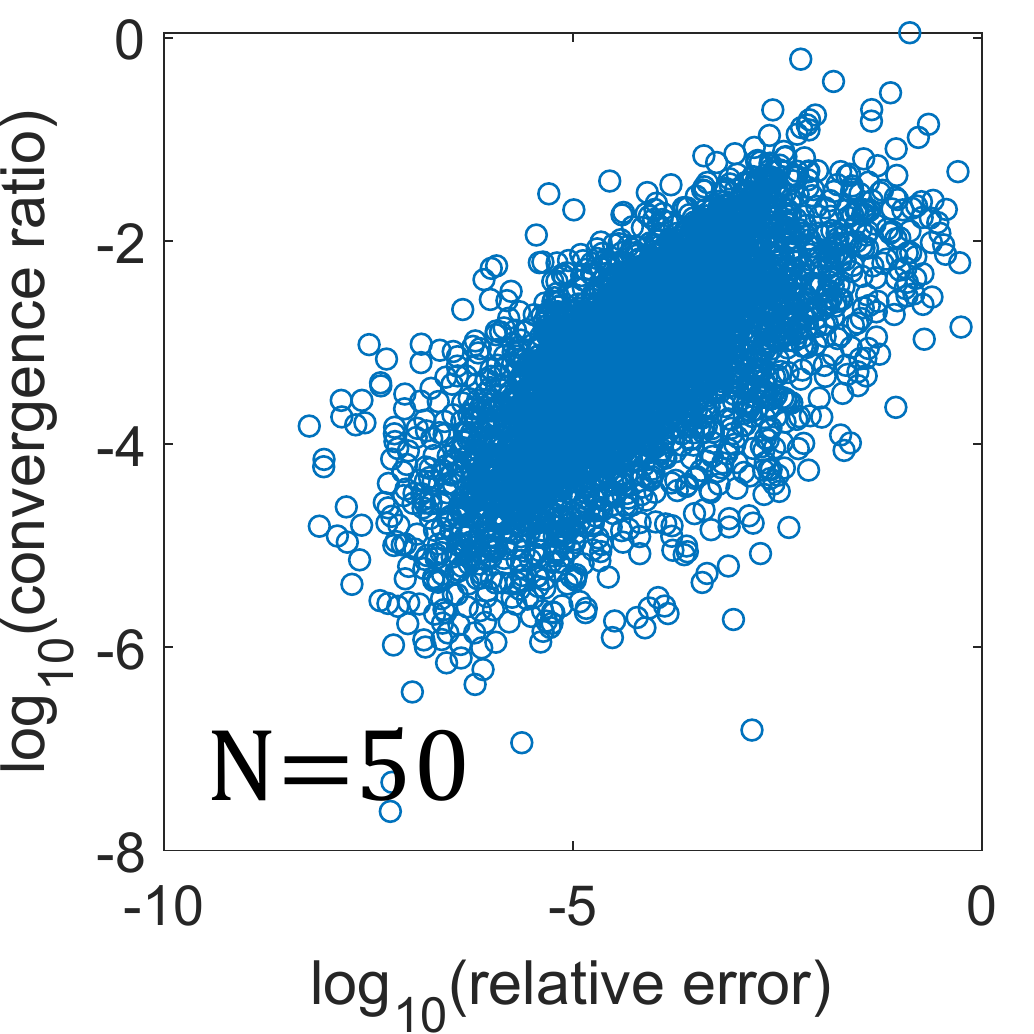}
    \end{subfigure}
    \caption{{\bf Convergence to true extinction time.} {\bf Left:} Average relative error, $G_m$, for initial ecosystem sizes of $N=20$ ({\bf top}) and $N=50$ ({\bf bottom}) with interaction density $\rho=[0.3,0.5,1.0]$ and scaling parameter $\mu=[10^{-4},10^{-3},\dots,1]$ for ${\bf A}(t)={\bf A_0}+{\bf H_1}t$. Each line represents one parameter set $(\rho,\mu)$ averaged across $5000$ random ecosystems. The zeroth-order starts the iteration with $\tau_0=0$ so it always starts with 100\% error. {\bf Right:} Convergence ratio estimated using Eqn.\ref{eq:relaxation_v2_convergence_final} for the $m=10$ iteration compared to $G_{10}$. Generally higher relative errors in $\tau$ positively correlates with high high values of the convergence ratio.}
    \label{fig:convergence}
\end{figure}

We tested the convergence of this scheme for $N=20$ and $N=50$ by varying the parameters used to generate an adaptive dynamics given by ${\bf A}(t)={\bf A_0}+{\bf H_1}t$ for interaction densities $\rho=[0.3,0.5,1.0]$ and a wide range of random constant ${\bf H_1}$ with different rate scales $\mu=[10^{-4},10^{-3},\dots,1]$. Fig.\ref{fig:convergence} shows how well this scheme convergences to the true extinction time, $\tau$, given by numerically evaluating Eqn.\ref{eq:const_eigen} based on starting with the zeroth-order or the second-order estimate of the earliest extinction time. We then compare the relative error, $G_m=(\tau-\tau_m)/\tau$, for each iteration $m$. This result shows good convergence even when we start with $\tau_0=0$.

However, there are several limitations involved in this calculation. First, there is no guarantee that any species will go extinct. For small ecosystems it is possible that all returned coefficients are positive such that all species are growing to infinity. Second, because we are dealing with only a single expansion around the initial starting time this requires that extinction must occur within the domain of convergence of our polynomial series given in Eqn.\ref{x(t)}. This is in general not true, especially for more complex dynamics involving higher-orders of ${\bf H_m}$. Overall, the best method to determine the extinct time involves iterating our full scheme as shown in Fig.\ref{fig:CLG_fig} when neither of the above conditions are reliably satisfied.

Unfortunately, there is no way to mathematically isolate this condition. However, we have shown that when these conditions are meant, our scheme provides a good estimate of the extinction time. For example, truncating the polynomial series at second-order can reasonably estimate the extinction time within 1-10\% error on average.

\section{Species abundances for the special case of ${\bf A}={\bf A_0}+{\bf H_1}t$}
\label{app:constant_exact}
For an initial $N$-species community described by $({\bf A},\vec{r})$ and the mean-field approximation to the changes of the interactions ${\bf H}t$ we have that the fixed-point of the system at some time $t$ can be calculated as $\vec{x}(t) = - ({\bf A} + {\bf H} t)^{-1}\vec{r} = ({\bf I} + {\bf A}^{-1}{\bf H}t)\vec{x}_0$ where $\vec{x}_0$ is the initial fixed-point of the system. Using the implicit form $({\bf I}+ {\bf A}^{-1}{\bf H}t)\vec{x}(t) = \vec{x}_0$ and if the initial and time-dependent fixed-points can be written as a linear combination of the projections on the eigenvectors of the operator ${\bf U}=-{\bf A}^{-1}{\bf H}$ we then have that
\begin{equation*}
\sum_{j=1}^N \left(1-\lambda_j t\right)c_j(t) \vec{v}_j = \sum_{j=1}^N c_j(0) \vec{v}_j
\end{equation*}
where $\vec{x}_0 = \sum_{j=1}^N c_j(0) \vec{v}_j$, $\vec{x}(t) = \sum_{j=1}^N c_j(t) \vec{v}_j$ and ${\bf U}\vec{v}_j = \lambda_j \vec{v}_j$. If the eigenvectors are linearly independent and normalized, we can solve for $c_j(t)$ to get
\begin{equation*}
c_j(t) = \frac{c_j(0)}{1-\lambda_j t}.
\end{equation*}

Finally, this allows us to write down the fixed-point at some time $t$ to be
\begin{equation*}
\vec{x}(t) = \sum_{j=1}^N \frac{c_j(0)}{1-\lambda_j t} \vec{v}_j.
\end{equation*}
Note that $c_j(0)$ is the length of $\vec{x}_0$ along the $j^\mathrm{th}$ eigenvector. That is, $\vec{c}(0) = {\bf V^{-1}}\vec{x}_0$, where ${\bf V}$ is a matrix whose $j^\mathrm{th}$ column is the $j^\mathrm{th}$ eigenvector $\vec{v}_j$.

The possible extinction time, $\tau_i$, for the $i$-th species is the earliest real, positive root of the $N$-order polynomial, $x_i(\tau_i)=0$, and if this root does not exist then the $i$-th species does not go extinct in the current community. The \emph{actual} extinction of the current community, $\tau=\min_i \tau_i$, is the earliest extinction time of all the species and all later possible extinction times are invalid. This is because as soon as a species dies the community reduces in size and the operator ${\bf U}$ also changes. The additional constraints on the applicability of this result can be found in Appendix \ref{app:limits_of_theory}.

\section{Time to extinction for the special case of ${\bf A}={\bf A_0}+{\bf H_1}t$}
    \label{app:relaxation_scheme_constant_exact}

In this section we will approximate the extinction time for communities which have an interaction matrix changing as ${\bf A}(t)={\bf A}_0 + {\bf H}t$. Although the exact solution is given by Eqn.\ref{eq:const_eigen} it is non-trivial to calculate the extinction time. For a community composed of $N$ species we must solve a system of $N$ equations where each equation is the root of an $N$-order polynomial with respect to time $t$. Then filter the resulting roots for a realistic extinction time. Additionally, we hope that writing down the expression for the extinction time will provide some insights in how community parameters such as interaction density, interaction strength, etc. can alter the rate of collapse for the community. 

We will apply a perturbation scheme on the exact solution given by Eqn.\ref{exactxwithU1} to approximate the extinction time of the community. First setting Eqn.\ref{exactxwithU1} to zero to find the extinction time $\tau$
\begin{equation*}
0=\vec{x}_0+{\bf U_1}\tau({\bf I}-{\bf U_1}\tau)^{-1}\vec{x}_0
\end{equation*}
gives us a system of equations used to determine the extinction time for each species. However, only the first species which goes extinct will realistically do so. The other species will rearrange themselves into a smaller community which we must solve separately to determine the next extinction time.

We apply a perturbation scheme to estimate the extinction time of the community. First, we substitute $\tau_m$ and $\tau_{m-1}$ for the two $\tau$'s, 
\begin{equation}
    0 = \vec{x}_0+{\bf U_1}\tau_m({\bf I} -{\bf U_1}\tau_{m-1})^{-1}\vec{x}_0.
    \label{eq:app_roots_with_U1}
\end{equation}
Then by repeatedly substituting in our guess of the extinction time into $\tau_{m-1}$ we can solve for a more accurate extinction time $\tau_m$. Using the first guess $\tau_0 = 0$ we can find the extinction time $(m=1)$ for the $i^\mathrm{th}$ species
\begin{equation}
    \tau_{1i} = \frac{-[\vec{x}_0]_i}{[{\bf U_1}\vec{x}_0]_i}.
\end{equation}
This term approximates the behavior of $\vec{x}(t)$ near $t=0$. If the species has a positive slope then the extinction time returned will be \emph{negative} and if it appears that the population abundance is declining then the extinction time will be \emph{positive}. However, the species with the earliest positive extinction time will die first which will cause the community to rearrange in its absence. Therefore, we must examine each extinction time, discarding the negative extinction times, and select only the earliest extinction time. We perform this positive minimization and define it as the first-order approximation, $\tau_1=\min_i\tau_{1i}$. This minimization also specifies which species is going extinct in the approximation.

We can now plug this estimate into Eqn.\ref{eq:app_roots_with_U1} and solve for the next estimate $(m=2)$
\begin{equation}
    \tau_{2i} = \frac{-[\vec{x}_0]_i}{[{\bf U_1}({\bf I}-{\bf U_1}\tau_1)^{-1}\vec{x}_0]_i}
\end{equation}
This will again give $N$ possible extinction times from which we discard the negative times and pick the earliest positive time, $\tau_2 = \min_i\tau_{2i}$. This process can be repeated $m$ times to find
\begin{equation}
    \tau_{mi} =  \frac{-[\vec{x}_0]_i}{[{\bf U_1}({\bf I}-{\bf U_1}\tau_{m-1})^{-1}\vec{x}_0]_i}
    \label{eq:app_tau_m_with_U1}
\end{equation}
where a positive minimization is required after each step and tells us which species is likely to die.

If the scheme converges to the true extinction time $[\vec{x}]_i(\tau)=0$ for the extinction of species $i$ then we would expect the error between each iteration to decrease in magnitude. We now ask how does the error, $\delta_m$, in the $m^\mathrm{th}$ estimate change as $m$ increases.

First, we shift to the basis which diagonalizes ${\bf U_1}$. By defining the $i^\mathrm{th}$ eigenvalues $\lambda_i$ and normalized eigenvector $\vec{v}_i$ of ${\bf U_1}$ as
\begin{equation*}
    {\bf U_1}\vec{v}_i = \lambda_i\vec{v}_i
\end{equation*}
we can find the projection of $\vec{x}_0$ on this basis as
\begin{equation*}
    \vec{x}_0 = \sum_{k=1}^N c_k\vec{v}_k
\end{equation*}
where the projection coefficients are given by
\begin{equation*}
    \vec{c} = {\bf V^{-1}}\vec{x}_0
\end{equation*}
and ${\bf V}$ is the matrix of eigenvectors with $\vec{v}_i$ in its $i^\mathrm{th}$ column.

Then, substituting the error terms into Eqn.\ref{eq:app_tau_m_with_U1} in the new basis we have
\begin{align}
    \tau + \delta_m =  \frac{-[\vec{x}_0]_i}{\sum_{k=1}^N\frac{\lambda_k}{1-\lambda_k(\tau +\delta_{m-1})} c_k [\vec{v}_k]_i}
\end{align}
where the term in the denominator is approximated as
\begin{align}
    \frac{\lambda_k}{1-\lambda_k(\tau+\delta_{m-1})} &=\frac{\lambda_k}{(1-\lambda_k\tau)(1-\frac{\lambda_k\delta_{m-1}}{1-\lambda_k\tau})}\nonumber\\
    &\approx\frac{\lambda_k}{1-\lambda_k\tau}\left(1+\frac{\lambda_k\delta_{m-1}}{1-\lambda_k\tau}\right)
\end{align}
assuming that $|\lambda_k \delta_{m-1}| \ll |1-\lambda_k\tau|$. If we apply this assumption to all eigenvalues in the summation then we have must have $\delta_{m-1}\ll|1-\lambda_{\max}\tau|/|\lambda_{\max}|$ where $\lambda_{\max}$ is the eigenvalue with the largest magnitude. If our iteration has a small enough error to the true extinction time we can then find
\begin{equation}
    \tau + \delta_m \approx \frac{-[\vec{x}_0]_i}{\sum_{k=1}^N\frac{\lambda_k}{1-\lambda_k\tau}\left(1+\frac{\lambda_k\delta_{m-1}}{1-\lambda_k\tau}\right) c_k [\vec{v}_k]_i}.
\end{equation}
Using the definition for the extinction time
\begin{equation*}
    0=[\vec{x}_0]_i+\tau\sum_{k=1}^N\frac{\lambda_k}{1-\lambda_k\tau}c_k[\vec{v}_k]_i
\end{equation*}
we can simplify the denominator to
\begin{align}
    \tau + \delta_m &= \frac{-[\vec{x}_0]_i}{\frac{-[\vec{x}_0]_i}{\tau}+\delta_{m-1}\sum_{k=1}^N\frac{\lambda_k}{1-\lambda_k\tau}\left(\frac{\lambda_k}{1-\lambda_k\tau}\right) c_k [\vec{v}_k]_i}\nonumber\\
    &=\frac{\tau}{1-\delta_{m-1} z}
\end{align}
where $z =\frac{\tau}{[\vec{x}_0]_i}\sum_{k=1}^N\frac{\lambda_k}{1-\lambda_k\tau}\left(\frac{\lambda_k}{1-\lambda_k\tau}\right) c_k [\vec{v}_k]_i$ depends upon the true extinction time of the $i^\mathrm{th}$ species. Assuming that $|\delta_{m-1}|\ll|z|$ we finally arrive at
\begin{equation}
    \delta_m \approx \delta_{m-1}\tau z
\end{equation}
so the ratio between the new and old errors
\begin{equation*}
    \left|\frac{\delta_m}{\delta_{m-1}}\right|=\left|\frac{\tau^2}{[\vec{x}_0]_i}\sum_{k=1}^N\frac{\lambda_k}{1-\lambda_k\tau}\left(\frac{\lambda_k}{1-\lambda_k\tau}\right) c_k [\vec{v}_k]_i\right|
\end{equation*}
will decrease as long as this ratio is less than unity. Although there are $N$ possible extinction times to examine in each iteration, we always pick the earliest positive extinction time so there should only be one associated time ``error" for the convergence.
\section{Canonical adaptive dynamics}
\label{app:dAdt_and_H}
Here we derive the {\bf H} matrices for the special case of mutation driven AD using the canonical form of AD. We assume that an interaction function has been chosen and that the derivatives of each interaction with respect to their dependent traits are known. We then use the framework of AD to construct the each order of ${\bf H}$ shown in Eqn.\ref{eq:A_Ht}.

In standard adaptive dynamics for each species we can assign a number of traits $S_k$ for the $k^\mathrm{th}$ species and calculate the total rate of change for the interactions based on all possible changes in the traits. Let $s_{k\alpha}$ denote the $\alpha^\mathrm{th}$ trait of species $k$ and sum over all possible changes for the interaction $A_{ij}$ to get
\begin{equation*}
    \frac{dA_{ij}}{dt} = \sum_{k=1}^N\sum_{\alpha=1}^{S_k} \frac{\partial A_{ij}}{\partial s_{k\alpha}}\frac{ds_{k\alpha}}{dt}
\end{equation*}
where the rate of change for each traits, $s_{k\alpha}$, is given by \cite{dieckmann1996dynamical} as
\begin{equation*}
    \frac{ds_{k\alpha}}{dt} = \frac{1}{2}\mu_{k\alpha} \sigma_{k\alpha}^2 \frac{x_k}{\lambda} \frac{\partial f_k}{\partial s_{k\alpha}}
\end{equation*}
where $s_{k\alpha}$ is the $\alpha^\mathrm{th}$ trait of species $k$, $\mu_{k\alpha}$ and $\sigma^2_{k\alpha}$ are the mutation rate and variance of mutation strength for $s_{k\alpha}$, $\lambda$ is the abundance corresponding to a single individual so that $x_k/\lambda$ represents the number of individuals, and $f_k$ is the LV fitness given by
$f_k = r_k + \sum_l A_{kl}x_l.$
The interspecies interactions $A_{ij}(\vec{s}_i,\vec{s}_j,\ldots)$ are functions of some traits of species $i$, $j$, or more.

We now evaluate the slope of the fitness hill
\begin{equation}
    \frac{\partial f_k}{\partial s_{k\alpha}} = \sum_{q=1}^N\left[ \frac{\partial A_{kq}}{\partial s_{k\alpha}} x_q+A_{kq}\frac{\partial x_q}{\partial s_{k\alpha}}\right].\label{chain}
\end{equation}
Note that $f_k$ is the relative fitness of a new mutant as it first emerges in a wild-type population where every other individual is $s$. The first term quantifies the relative fitness of a mutant individual that interacts differently with others, while the second term quantifies the relative fitness of the newly emerged mutant due to a change it causes to the other species' abundances. Since the new mutant is very few in abundance compared to the wild type, the abundances of others will not depend on its trait value, and thus $dx_q/ds_{k\alpha}=0$. To understand this point better with specific examples, the reader could refer to \cite{dieckmann1996dynamical,marrow1996evolutionary} (in their notation $dx/ds'=0$). Therefore, we have

\begin{equation*}
    \frac{ds_{k\alpha}}{dt} = \frac{1}{2}\mu_{k\alpha}\sigma_{k\alpha}^2\frac{x_k}{\lambda}\sum_{q=1}^N\frac{\partial A_{kq}}{\partial s_{k\alpha}}x_q.
\end{equation*}

Finally, we can write down the rate of change for each pair of interactions as
\begin{equation}
    \frac{dA_{ij}}{dt} = \frac{1}{2\lambda}\sum_{k=1}^N  x_k \sum_{\alpha=1}^{S_k}\mu_{k\alpha}\sigma_{k\alpha}^2\frac{\partial A_{ij}}{\partial s_{k\alpha}}\sum_{q=1}^N \frac{\partial A_{kq}}{\partial s_{k\alpha}}x_q.
    \label{eq:dAdt}
\end{equation}

The interaction derivatives can be any general function of time, $t$. To write $\frac{d{\bf A}}{dt}$ as shown in Eqn.\ref{eq:A_Ht} we must now assume that each derivative is approximately constant
\begin{equation*}
    \frac{\partial A_{ij}}{\partial s_{k\alpha}} = C_{ijk\alpha}.
\end{equation*}
Although we take this simplifying assumption, it is not completely necessary. As long as $\frac{\partial A_{ij}}{\partial s_j}$ can be written as a power series of $t$ then this method will still work but will require updating to fully match the constraints of Eqn.\ref{eq:A_Ht}. A fully general prescription for this derivation is outside the scope of this Appendix.

We now seek to break down, term by term, the time dependencies of each expression shown in Eqn.\ref{eq:dAdt}. First, using $[\vec{x}]_q = [\vec{\epsilon}_0]_{q} + [\vec{\epsilon}_1]_{q}t + [\vec{\epsilon}_2]_{q}t^2+\dots$ we have
\begin{equation*}
    \sum_{q=1}^N C_{kqk\alpha} [\vec{x}]_q = \sum_{q=1}^N C_{kqk\alpha}([\vec{\epsilon}_0]_{q} + [\vec{\epsilon}_1]_{q}t + [\vec{\epsilon}_2]_{q}t^2+\dots).
\end{equation*}
Then
\begin{align*}
    x_k&\sum_{q=1}^N C_{kqk\alpha}x_q\\
    =&\sum_{q=1}^N C_{kqk\alpha}([\vec{\epsilon}_0]_{k}+[\vec{\epsilon}_1]_{k}t+\dots)([\vec{\epsilon}_0]_{q}+[\vec{\epsilon}_1]_{q}t+\dots) \\
    =&\sum_{m=0}^\infty \left(\sum_{s=0}^m [\vec{\epsilon}_s]_{k}\sum_{q=1}^N C_{kqk\alpha}[\vec{\epsilon}_{m-s}]_{q}\right)t^m
\end{align*}
We then arrive at
\begin{align*}
    \frac{dA_{ij}}{dt} = \sum_{m=0}^\infty\sum_{k=1}^N \sum_{\alpha=1}^{S_k}\frac{\mu_{k\alpha}\sigma_{k\alpha}^2}{2\lambda} C_{ijk\alpha}\times\\
    \left(\sum_{s=0}^m [\vec{\epsilon}_s]_{k}\sum_{q=1}^N C_{kqk\alpha}[\vec{\epsilon}_{m-s}]_{q}\right)t^m.
\end{align*}
This arranges the summation with respect to their power of $t$ for $\frac{d{\bf A}}{dt}$. To find the expression for ${\bf A}(t)$ we finally \emph{integrate} over time to find the $m$-order of ${\bf H_m}$ ($m>0$) in Eqn.\ref{eq:A_Ht} as
\begin{equation}
    [{\bf H_m}]_{ij} = \sum_{q,k=1}^N\sum_{\alpha=1}^{S_k}\frac{\mu_{k\alpha}\sigma_{k\alpha}^2}{2m \lambda} C_{ijk\alpha} \sum_{s=0}^{m-1} [\vec{\epsilon}_s]_{k} C_{kqk\alpha}[\vec{\epsilon}_{m-s-1}]_{q}.
    \label{eq:H_m}
\end{equation}
Then using $\vec{x}(t)$ generated we can then predict the behavior of the traits by integrating $\frac{d\vec{s}}{dt}$ over time. Given the initial trait values as $\vec{s}_0 = \vec{s}(t=0)$ we then have
\begin{align}
    s_{k\beta}(t) &= s_{k\beta}(0) + \frac{\vec{\mu}_{k\beta}\sigma_{k\beta}^2}{2\lambda}\sum_{\alpha=1}^{S_{k}}\bigg[[\vec{x}_0]_k\sum_{q=1}^N C_{kqk\alpha}[\vec{x}_0]_q t\nonumber\\
    &+\bigg([\vec{x}_0]_k\sum_{q=1}^N C_{kqk\alpha}[\vec{\epsilon}_1]_q+[\vec{\epsilon}_1]_k\sum_{q=1}^N C_{kqk\alpha}[\vec{x}_0]_q\bigg) \frac{t^2}{2}\nonumber\\
    &+\dots\bigg]
\end{align}
where $s_{k\beta}$ is the $\beta$-th trait for species $k$. 

\section{Technical discussion on the convergence and rate of convergence of series solutions}
\label{app:limits_of_theory}
We now demonstrate why repeated iterations of perturbation theory are required in order to examine the abundances for longer times. Consider the constant rate model ${\bf A}(t) = {\bf A_0}+{\bf H_1}t$ for which we have both the analytical solution using the diagonal basis of the operator ${\bf U} = -{\bf A_0^{-1}H_1}$ and the perturbation scheme described in the previous section. We now use the results from the perturbation theory to derive the analytic solution. First, note that ${\bf U}$ is also used to get the coefficients in the perturbation theory. Then by projecting the initial population abundance vector into the eigenvectors of ${\bf U}$ we have that $\vec{\epsilon}_0 = \sum_j c_j(0) \vec{v}_j$ where ${\bf U}\vec{v_j} =\lambda_j \vec{v}_j$. Therefore the polynomial expansion from the previous section can be written as
\begin{align*}
    \vec{x}(t) =& \sum_j(1 + \lambda_j t + \lambda_j^2 t^2 + \dots)c_j(0) \vec{v}_j\\
    =& \sum_j c_j(0) \vec{v}_j \sum_{m=0}^\infty (\lambda_j t)^m
    = \sum_j \frac{c_j(0) \vec{v}_j}{1-\lambda_j t}
\end{align*}
which is the same result as the analytic method. However, there are two restrictions for the valid time domain of this solution. First, the solution can be divergent for certain values of $\lambda$. Second, when condensing the geometric series (polynomial expansion) from perturbation theory to the analytic expression we have implicitly assumed that we are within the convergence region of the geometric series. Therefore the spectral radius, $|\lambda|_{\max}$, of the operator ${\bf U}$ restricts the convergent domain of $t<\frac{1}{|\lambda|_{\max}}$.

The polynomial series from perturbation theory is more restrictive than the analytic result from using eigenvalues due to the convergence condition. Even if there are no singularities for positive (real) times the result from perturbation theory only converges for a time interval restricted by the magnitude of the largest eigenvalue. This restriction is always applied because we are approximating the analytic result with a series which in application is truncated. This restriction applies to any form of $\frac{d{\bf A}}{dt}$ because the perturbation theory approximates the operation of applying a matrix inversion continuously over time.

Typically the expansion will blow up to infinity for long times and can alternate between positive and negative infinity as the truncated order is increased. Therefore, we can use a heuristic estimate to determine a good time step as
\begin{equation}
    T=\left( \frac{\chi}{\max_i|[\vec{\epsilon}_m]_i|}\right)^{1/m}    
\end{equation}
where $\chi$ is the allowed error in the approximation for $\vec{x}(t)$ and assuming that a sufficiently accurate order $m$ has been chosen for $\vec{x}(t)=\vec{x}_0+\vec{\epsilon}_1t+\dots+\vec{\epsilon}_m t^m$. This is because the highest-order term serves as an estimation for the error in our approximation of $\vec{x}(t)$ when you compare two power series for $\vec{x}(t)$ of different truncation orders $m-1$ and $m$.

\end{document}